\pdfoutput=1

\documentclass{iopart}

\usepackage{epsfig}
\usepackage{amscd,amssymb}
\usepackage{graphicx,xspace}
\usepackage{pstricks}
\usepackage[left=3cm,top=3cm,right=3cm,bottom=3cm]{geometry}
\usepackage{color}

\begin{document}

\title{Sinai model in presence of dilute absorbers}

\author{ Pierre Le Doussal}
\address{CNRS-Laboratoire de Physique Th\'eorique de l'Ecole\\
Normale Sup\'erieure, 24 rue Lhomond, 75231 Paris
Cedex-France\thanks{LPTENS is a Unit\'e Propre du C.N.R.S.
associ\'ee \`a l'Ecole Normale Sup\'erieure et \`a l'Universit\'e Paris Sud}}

\begin{abstract}
We study the Sinai model for the diffusion of a particle in a one dimension random potential in presence of a 
small concentration $\rho$ of perfect absorbers using the asymptotically exact real space renormalization method. 
We compute the
survival probability, the averaged diffusion front and return probability, the two particle meeting probability, the distribution of total distance traveled before absorption and the averaged Green's function of the associated Schrodinger operator. Our work confirms some recent results of Texier and Hagendorf obtained by Dyson-Schmidt methods, and extends them to other observables and in presence of a drift. In particular the power law density of states is found to hold in all cases. Irrespective of the drift, the asymptotic rescaled diffusion front of surviving particles is found to be a symmetric step distribution, uniform for $|x| < \frac{1}{2} \xi(t)$, where $\xi(t)$ is a new, survival length scale ($\xi(t)=T \ln t/\sqrt{\rho}$ in the absence of drift). Survival outside this sharp region is found to decay with a larger exponent, continuously varying with the rescaled distance $x/\xi(t)$. A simple physical picture based on a saddle point is given, and universality is discussed. 
\end{abstract}


\maketitle



\section{Introduction and model}

The Sinai model \cite{sinai} of a particle undergoing thermally activated diffusion in a one dimensional random energy landscape $U(x)$ is the simplest tractable model of glassy dynamics and has generated considerable attention in physics, in probability theory and even in biophysics and in finance \cite{oldresults,biophysics,vinokur,finance}. In this model $U(x)$ has the statistics of an unbiased random walk, i.e. $\overline{(U(x)-U(x'))^2} \sim 2 \sigma |x-x'|$ hence the barriers grow as $\sqrt{|x|}$ with spatial separation and as a consequence diffusion is ultra-slow as $x \sim T^2 \ln^2 t$. Interestingly, the associated Fokker-Planck operator maps to the Schrodinger operator of a 1D supersymmetric quantum mechanics describing a class of particle-hole symmetric random hopping Hamiltonian. These have been studied using replica \cite{oldresults}, Dyson-Schmidt \cite{pastur,oldresults,texierdos} and supersymmetry methods \cite{balentsfisher}. They exhibit the Dyson singularity $N(E)\sim 1/(\ln E)^2$ of the (integrated) DOS and the exact (quasi-delocalized) eigenstate at $E=0$ and are 1D analogs of the heavily studied 2D chiral class which in two dimension exhibit delocalization phenomena. The Sinai model enjoys many remarkable properties, among them anomalous drift $x \sim t^\mu$ in presence of a bias in $U(x)$, followed by a transition to non-zero velocity \cite{Kesten}. Another remarkable property \cite{golosov}, at zero bias, is that, up to rare events, the thermal packet is concentrated in a finite region around the bottom of the potential well available at time $t$. This property has allowed to apply a powerful real space renormalization method (RSRG) to this problem and derive a number of exact results \cite{sinaiRSRG}. The idea is to eliminate iteratively all barriers smaller than $\Gamma=T \ln t$ and leads to results asymptotically exact in the large $\Gamma$ limit. This method was introduced in the closely related context of random quantum spin chains where it has proved extremely useful \cite{fisher1,igloimonthus}. 

Recently, Texier and Hagendorf (TH) have studied the Sinai model in presence of dilute perfect absorbers and found \cite{HT2} an interesting power law decay of the average probability of return to the origin, hence a power law behaviour of the density of states. This result was obtained using Dyson-Schmidt and related methods \cite{texierdos,HT1}. The aim of this paper is to study this problem using the RSRG method. Our results confirm some of the ones of TH and in addition we compute a number of other observables.

Let us consider the Sinai model, i.e a random walker performing Arrhenius diffusion at temperature $T$ in the landscape $U(x)=V(x)- f x$, i.e we allow for a bias towards the right $f>0$ and define $\delta =  f/(2 \sigma) >0$. We set $\overline{(V(x)-V(x'))^2} \sim 2 |x-x'|$, hence choose units of length such that $\sigma=1$. We do not here redefine the various realizations (discrete or continuum time and space) of the diffusion model, and we refer for that to Section II of Ref. \cite{sinaiRSRG}, our parameters being identical. If we now add a Poissonian absorption potential $W(x)$ with infinite strength, i.e a Poisson distribution of perfectly absorbing obstacles, the system is in effect cut into independent segments, i.e intervals between scatterers. Let us consider a model with impurity density $\rho$ and distribution of the length of segments between impurities is $p(L)=\rho e^{- \rho L}$. We are studying the dilute limit where the average distance $L_\rho=1/\rho$ is much larger than all other crossover or microscopic lengths. We assume no global constraint on the energy landscape $U(x)$, i.e. it performs a free random walk (i.e. in the universality class of the free Brownian motion at large scale, upon rescaling). In that case the landscape in each segment is also a free random walk (which assumes, however, the same value at the common boundaries of each pair of consecutive segments).

\begin{figure}[t]
  \centering
  \includegraphics[width=0.6\textwidth]{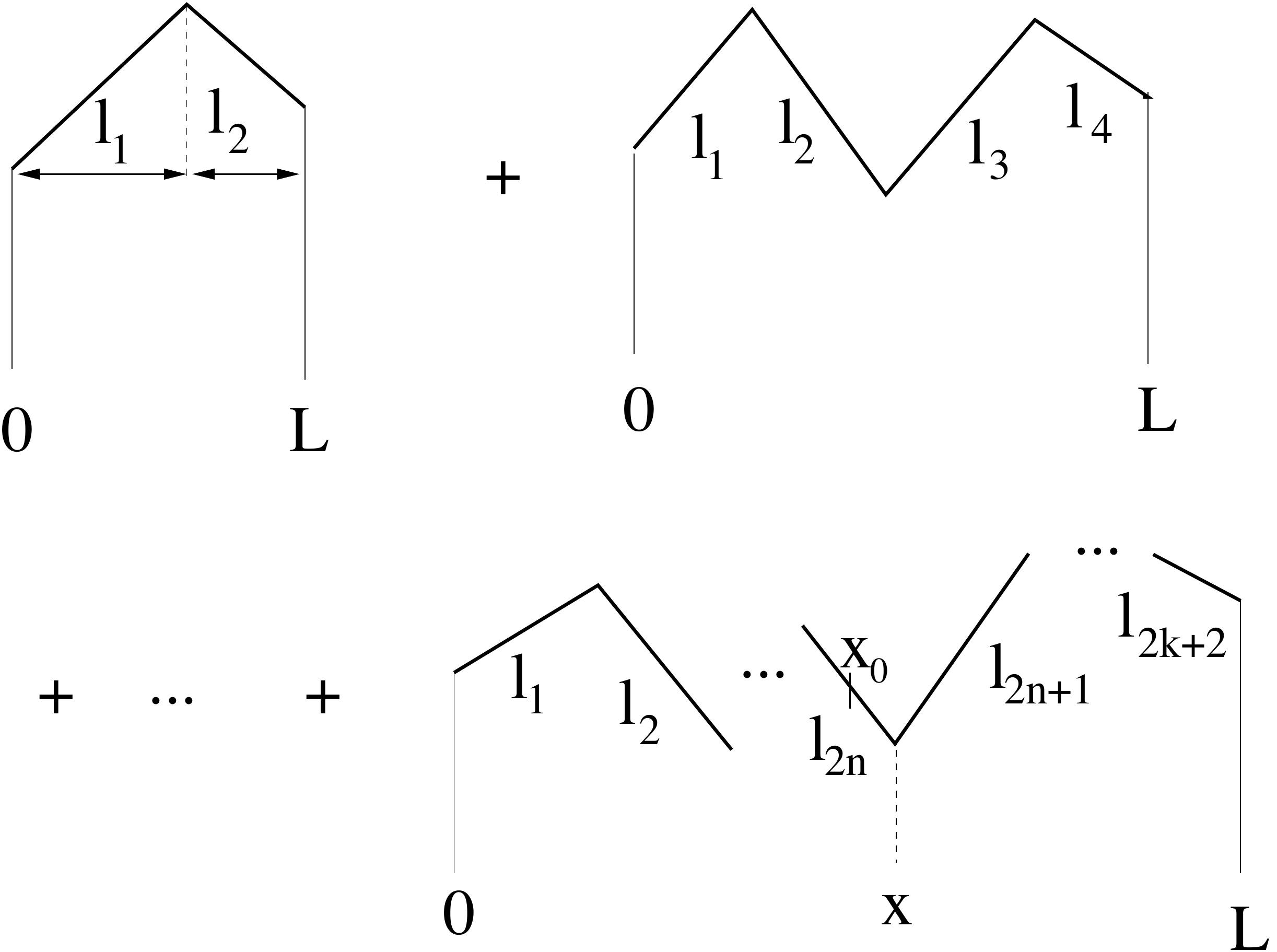}
 \caption{Schematic representation of the finite size RSRG measure \ref{fs} for the renormalized landscape ($x$ axis is length, the $\ell_i$ are the bond length, and the vertical $y$ axis - not drawn - is the disorder potential energy). It is a sum of 
 terms with $k=1,2,..$ valleys (i.e. sets of two consecutive bonds) not yet decimated at scale $\Gamma=T \ln t$, i.e. of barriers larger than $\Gamma$ (in vertical $y$ direction). The edge bonds (first and last) cannot be decimated, only the bulk bonds (all the others) are. In the absorber model, there is an absorber at $x=0$ and $x=L$. A particle which started at $t=0$ on the left edge bond (resp. right) has been absorbed at time $t$ by the left absorber at $x=0$ (resp. the right absorber at $x=L$), asymptotically with probability one. A particle which started at $x_0$ on bond $2 n$, $1 \geq n \geq k$ has survived and moved to the right at $x$, the bottom of the starting bond (asymptotically with probability one), while a particle starting on bond $2n+1$ - not shown - has moved to the left to the same point $x$ (asymptotically with probability one).}
    \label{fig:1}
\end{figure}

The outline of the paper is as follows. In Section \ref{sec2} we recall the RSRG method for the standard Sinai model in system of finite size $L$ with absorbing boundaries. In Section \ref{sec3} we average over the distances between absorbers and obtain the survival probability and its power law decay in time. In Section \ref{sec4} we compute the full averaged diffusion front and analyze its form at large time. The survival length scale $\xi(t)$ is introduced. In Section \ref{sec5} we obtain the average return probability. In Section \ref{sec6} we focus on the distribution of the relaxation rates, and obtain a simple physical picture of the power law in time relaxation in this model. We also analyze some rare events, and whether they could lead to slower decay. In Section \ref{sec7} we consider the diffusion of two particle and their meeting time. In Section \ref{sec8} we compute the distribution of distance traveled before absorption. In Section \ref{sec9} we obtain the Schrodinger Green's function, and a discussion and conclusion is given in Section \ref{sec10}. Appendice A contains more on two particule survival and Appendix B on the diffusion front in the absence of a random potential.

\section{Finite size measure}
 \label{sec2}

\begin{figure}[t]
  \centering
  \includegraphics[width=0.3\textwidth]{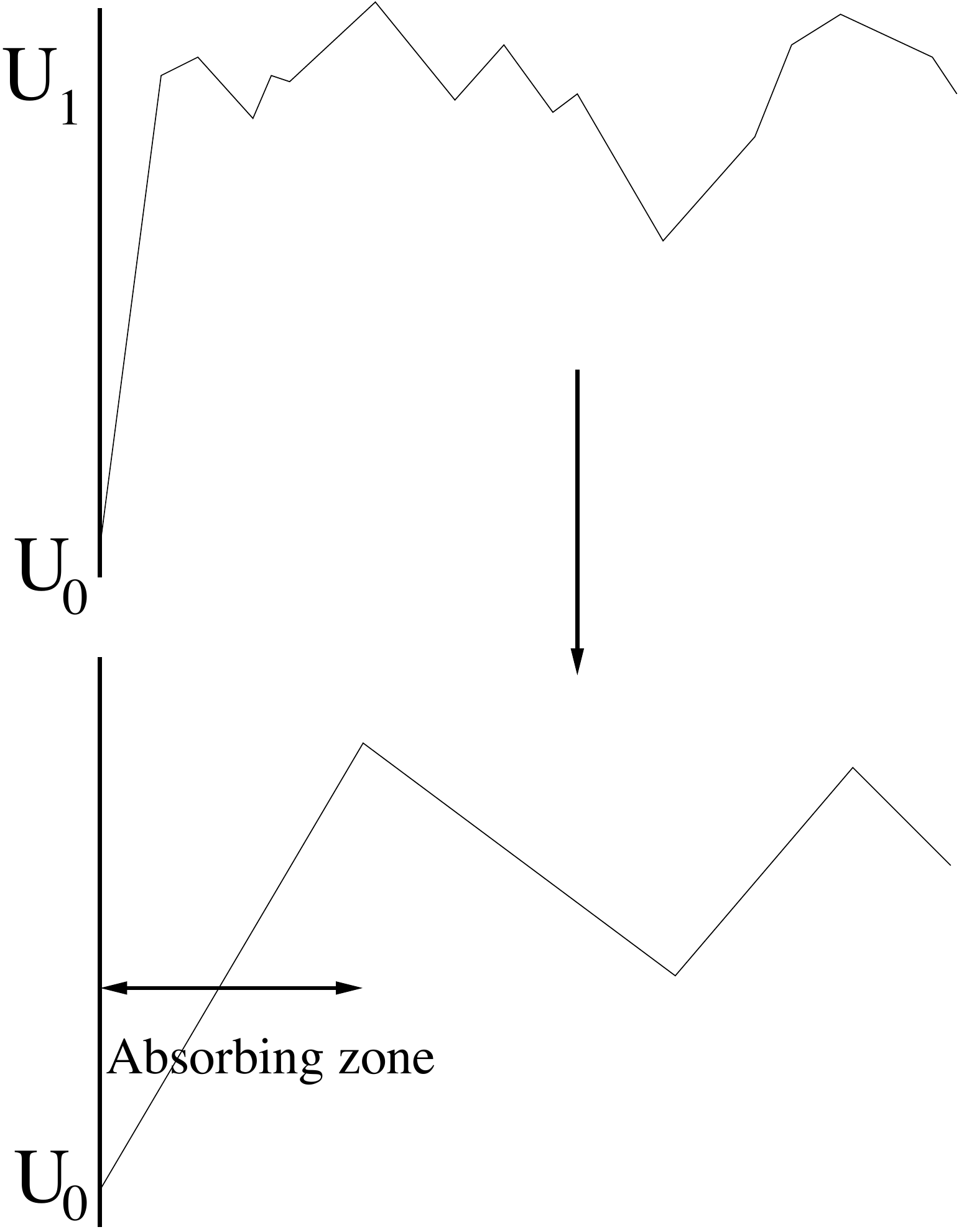}
 \caption{Absorbing boundary in the RSRG. Top: the initial landscape $U(x)$ for $x \geq 0$: $U_0$ is artificially set to a very large negative value. Bottom: the renormalized landscape after the iterative decimation of the smallest barriers up to barrier size $\Gamma$. The fist edge bond - which is never decimated - is an absorbing zone: any particle starting there at time $t=0$ has moved to the left edge while encountering only barriers smaller than $\Gamma$, hence it is absorbed with probability asymptotically equal to one at time $T \ln t = \Gamma$. Particles starting on the next (i.e. bulk) bond have moved to the right and have survived. Another such boundary is present at $x=L$ (not shown here).}
    \label{fig:2}
\end{figure}

Let us recall first the finite-size RSRG for the Sinai model. More details can be found in \cite{sinaiRSRG} and in e.g. \cite{fisheryoung,monthusFS} for quantum models. The measure for the renormalized landscape at scale $\Gamma$ for a system of fixed size $L$ can be expressed as a sum:
\begin{eqnarray} \label{fs} 
&& N_{\Gamma,L}[\ell_i] = \bar l_\Gamma E_\Gamma^-(\ell_1)E_\Gamma^+(\ell_2) \delta(L-(\ell_1+\ell_2)) \\
&& + \sum_{k=1}^{\infty} 
\bar l_\Gamma E_\Gamma^-(\ell_1) \big(
\prod_{j=1}^k P^+(\ell_{2 j}) P^-(\ell_{2 j+1})  \big) E_\Gamma^+(\ell_{2k+2} )  \delta(L- \sum_{i=1}^{2 k+2} \ell_i) 
\end{eqnarray}
i.e. a sum of measures for the events where there remain $2 k+2$ bonds in the system, i.e. $k$ valleys of bulk bonds and two edge bonds, with $k=0,1,..$. We have singled out the term $k=0$ which plays a special role. The product form reflects the Markovian nature of the landscape, which is preserved by decimation, hence is either an exact consequence of the choice of a Markovian initial landscape, or a consequence of the convergence to the fixed point landscape. The only constraint is the fixed total length, implemented by the delta functions. Bonds in the bulk have a length distribution $P^{\pm}_\Gamma(\ell)$ while edge bonds 
have length distribution $E^{\pm}_\Gamma(\ell)$. The subscript $+$ denote descending bonds along the bias, while $-$ are barriers which oppose the bias. As seen in Fig. 1 the left edge bond is ascending, and corresponds to the ''absorption zone'' of the left boundary, while the descending right edge corresponds to the ''absorption zone'' of the right boundary. This geometry correspond the finite size measure with so-called AA boundary conditions, i.e. absorbing on both ends as relevant for the present problem, also shown in Fig.  \ref{fig:2}. The factor $\bar l_\Gamma$ which is nothing but the average bond length ensures the normalization of the total probability to unity. Let us call $Z_L$ this probability. The measure is easier to write in Laplace transform with respect to $L$, i.e. multiplying (\ref{fs}) by $e^{- p L}$ and integrating $\int_0^\infty dL$ one gets the normalization condition $\int_0^\infty dL Z_L e^{-p L}=1/p$, i.e. $Z_L=1$, in the form:
\begin{eqnarray} \label{normal}
\bar l_\Gamma \frac{E_\Gamma^+(p) E_\Gamma^-(p)}{1 - P_\Gamma^+(p) P_\Gamma^-(p)} = \frac{1}{p} 
\end{eqnarray}
The fixed point form for the bond length probability, in Laplace $P_\Gamma(p)=\int_0^L e^{-p \ell} P_\Gamma(\ell)$, takes the form:
\begin{eqnarray} \label{formstart}
&& P_\Gamma^\pm(p) = \frac{\sqrt{p+\delta^2} e^{\mp \delta \Gamma}}{\sqrt{p+\delta^2} \cosh(\Gamma \sqrt{p + \delta^2}) \mp \delta \sinh(\Gamma \sqrt{p + \delta^2})} \\
&&  E_\Gamma^{\pm}(p) = \frac{\delta e^{\mp \delta \Gamma} }{\sinh(\delta \Gamma) ( \sqrt{p+\delta^2} \coth(\Gamma \sqrt{p + \delta^2}) \mp \delta)  }  \quad , \quad \bar l_\Gamma= (\frac{\sinh(\Gamma \delta)}{\delta})^2
\end{eqnarray}
which is easily checked to satisfy (\ref{normal}). 

Note the simpler form in the absence of bias:
\begin{eqnarray}
P_\Gamma(p) = \frac{1}{\cosh(\Gamma \sqrt{p})} \quad , \quad E_\Gamma(p) = \frac{\tanh(\Gamma \sqrt{p})}{\Gamma \sqrt{p}}  \quad , \quad \bar l_\Gamma= \Gamma^2
\end{eqnarray}
and the expressions after inverse Laplace transform (denoting for simplicity the function and its LT by the same symbol):
\begin{eqnarray} \label{direct}
&& P_\Gamma(\ell) = \frac{1}{\Gamma^2} \sum_{n=-\infty}^{+\infty} (n+\frac{1}{2}) \pi (-1)^n 
e^{- \frac{\pi^2}{\Gamma^2} (n+ \frac{1}{2})^2 \ell }  \quad , \quad 
 E_\Gamma(\ell)  = \frac{1}{\Gamma^2} \sum_{n=-\infty}^{+\infty} 
e^{- \frac{\pi^2}{\Gamma^2} (n+ \frac{1}{2})^2 \ell }  
\end{eqnarray}

We are now equipped to study the problem with absorbers. The idea is to use the effective dynamics: a particle starting at time $t=0$ at $x_0$ is  at time $t$ at the bottom of the bond (or the valley) containing $x_0$ in the renormalized landscape at $\Gamma=T \ln t$ \cite{sinaiRSRG}. It is there with a probability which tends to one as $\Gamma \to \infty$ (corrections due e.g. to barrier or well degeneracies, or anomalously rare thermal and disorder configurations can also be estimated). 
In the absence of absorbers this leads to a diffusion length scale given by the average bond length, $L(t)=l_\Gamma= \Gamma^2 = T^2 \ln^2 t$ in the symmetric case. Let us stress that while the RSRG is valid for any $\delta$ to determine
landscape extrema, its (simplest) application to the Arrhenius diffusion requires that the bias $\delta$ is small. Hence the results presented here will be valid to lowest leading order in $\delta$. Extensions to higher orders in $\delta$ are possible, but more difficult as the effective dynamics must then be modified to also take into account multiple well contributions order by order in $\delta$, see Ref. \cite{cecile}. Note finally that we denote disorder averages by $\overline{..}$ but use rather the notation $\langle .. \rangle_L$ for specific averages over the finite size block measure (\ref{fs}), although both have the same origin, i.e. the random potential $U(x)$ (we do not need here a notation for thermal averages).

\section{Survival probability} \label{survivalproba} 
\label{sec3}

Let us first compute the {\it survival probability} $S(t)$ of a walker up to time $t$. We assume that the walker starts at $x_0$ with uniform measure on space. Hence the probability density that it starts on a segment of length $L$ is $L p(L)/\int dl l p(l) = \rho L p(L)$. Given that, it has a uniform measure on the segment and its probability $S_L(\Gamma)$ to be alive at time $\Gamma= T \ln t$ is equal to the probability that $x_0$ is not in the absorbing zone, see Figs. \ref{fig:1} and \ref{fig:2}, equivalently that $x_0$ is on a bulk bond:
\begin{eqnarray}
S_L(\Gamma) = 1 -\frac{ \langle l_1 + l_{2k + 2} \rangle_L}{L} = \frac{\langle l_2+l_3+..+l_{2 k} \rangle_L}{L}
\end{eqnarray}

It is easy to compute the Laplace transform, using the normalization identity (\ref{normal}):
\begin{eqnarray}
\int dL L S_L(\Gamma) e^{- p L} = \frac{1}{p^2} +  \frac{E_\Gamma^{+ \prime}(p)}{p E_\Gamma^+(p)} + \frac{E_\Gamma^{- \prime}(p)}{p E_\Gamma^-(p)} 
\end{eqnarray}
Setting $p=\rho$ this immediately gives the survival probability in the original model, averaged over all segments. Let us first analyze the symmetric case:
\begin{eqnarray} \label{surv1} 
&& S(t) =  \rho^2 \int dL L S_L(\Gamma) e^{- \rho L} = \frac{2 \Gamma \sqrt{\rho}}{\sinh(2 \Gamma \sqrt{\rho})} 
 =  \frac{2 \sqrt{\rho} ~ T \ln t }{\sinh(2 \sqrt{\rho}  T \ln t)}  \\
 && = 1 - \frac{2}{3} \rho T^2 \ln^2 t + O((T \ln t)^4)  \quad , \quad T \ln t \ll 1/\sqrt{\rho} \\
&& \approx 4 \sqrt{\rho}  ~ (T \ln t)   ~  t^{- 2 T \sqrt{\rho}}  ~~~~~~~~~~~ \quad , \quad T \ln t \gg 1/\sqrt{\rho} \label{survlarget}
\end{eqnarray}
in respectively the small and large time limit. $S(t)$ correctly decreases from $1$ to zero as a power law at large time, plus logarithmic corrections. The crossover time occurs when the typical diffusion length in the absence of absorbers $L(t) = T^2 \ln^2 t $ becomes of the order of the typical separation between absorbers $L_\rho=1/\rho$. The validity of these results is small $\rho$ and large times, i.e. both lengths $L(t)$ and $L_\rho$ sufficiently large so that the renormalized landscape has reached its RSRG fixed point form. This implies that diffusion has crossed over from pure to ultra-slow (so short times here are always meant as large compared to the crossover time from pure diffusion to ultra-slow \cite{footnote1}). 

In presence of an applied force, i.e. a biased landscape one finds:
\begin{eqnarray} \label{survbias} 
&& S(t) =  \rho^2 \int dL L S_L(\Gamma) e^{- \rho L} =  2 \frac{\delta^2 + \rho T \ln t  \sqrt{\rho +\delta^2}  \coth(T \ln t \sqrt{\rho+\delta^2}) }{ 2 \delta^2 + \rho + \rho \cosh(2 T \ln t \sqrt{\rho+\delta^2}) } 
\end{eqnarray}
a generalization of the previous result. The general validity of this result is now small $\rho \sim \delta^2$. Let us recall that in the absence of absorbers there is a crossover length scale $L_\delta = 1/\delta^2$ below which the diffusion remains ultraslow and almost unaffected by the bias. Hence it is not surprising that the condition for formula (\ref{survbias}) to crossover to the result for the symmetric case (\ref{surv1}) is $\delta^2 \ll \rho$, i.e. $L_\rho \ll L_\delta$. 

We can now define a small time regime $T \ln t \ll 1/\sqrt{\rho + \delta^2}$ where:
\begin{eqnarray} 
&& S(t) = 1 - \frac{2}{3} \rho T^2 \ln^2 t - \frac{2}{45} \rho ( 8 \delta^2 - 7 \rho) T^4 \ln^4 t + O((T \ln t)^6)  
\end{eqnarray}
and a large time regime $T \ln t \gg 1/\sqrt{\rho + \delta^2}$ where the formula (\ref{survbias}) can be approximated by:
\begin{eqnarray}  \label{approx1} 
&& S(t) \approx 4 \frac{(\delta^2/\rho) + T \ln t  \sqrt{\rho +\delta^2}  }{ 4 (\delta^2/\rho) + t^{2 T \sqrt{\rho+\delta^2}}} \quad , \quad T \ln t \gg 1/\sqrt{\rho+\delta^2} 
\end{eqnarray}
In the general case $\rho \sim \delta^2$ this yields the large time decay of the survival probability:
\begin{eqnarray} 
&& S(t) \approx 4 ( \frac{ \delta^2}{\rho}  +  T \ln t  \sqrt{\rho +\delta^2} ) ~ t^{- 2 T \sqrt{\rho+\delta^2}} \quad , \quad T \ln t \gg 1/\sqrt{\rho+\delta^2} \quad , \quad \rho \sim \delta^2
  \label{survbiaslarget}
\end{eqnarray}
a simple generalization of the symmetric result (\ref{survlarget}). In the case where $\rho \ll \delta^2$, i.e. $L_\rho \gg L_\delta$, there is however another crossover time in the large time regime. As clear from (\ref{approx1}), it is defined by $1 \ll \sqrt{\rho+\delta^2} T \ln t \sim 4 \ln(\delta^2/\rho)$, i.e. time around $t \sim (\delta^2/\rho)^{4/\delta T}$ in which case:
\begin{eqnarray} 
&& S(t) \approx 1/(1 + \frac{\rho}{4 \delta^2} t^{2 T \sqrt{\rho+\delta^2}} )
\end{eqnarray} 
crosses over from $S(t)=1$ to a small $S(t)$. This additional crossover time is necessary since one must recover a unit survival probability when $\rho \to 0$, and that is not obvious on (\ref{survbiaslarget}). Hence when $\rho \ll \delta^2$, i.e. $L_\rho \gg L_\delta$, the large time regime splits in two subregimes depending on whether (i) $1 \ll \sqrt{\rho+\delta^2} T \ln t \ll 4 \ln(\delta^2/\rho)$ in which case $S(t) \approx 1$ and (ii) $\sqrt{\rho+\delta^2} T \ln t \gg 4 \ln(\delta^2/\rho)$ in which case $S(t)$ is again given by (\ref{survbiaslarget}).

The decay is thus faster in presence of a bias. This is intuitively clear since in the absence of absorbers the motion of the particle is faster at large time and becomes $\overline{x(t)} \sim t^{2 \delta T}/(4 \delta^2)$: hence the particle should die sooner. This effect 
however does not make the decay of the average survival probability exponential (or stretched exponential as in the absence of the random potential), here it remains a power law in time. This is because of the rare regions associated to a barrier against the bias which contain particles trapped for a long time. As found in \cite{sinaiRSRG} in presence of a small bias $\delta \ll 1$ and without absorbers, the diffusion front at large time takes the form $\overline{P(x,t|0,0)} \approx \theta(x)  \exp(-x/\overline{x(t)} )/\overline{x(t)}$ hence the probability that a particle has remained in a region of size $\sim 1/\rho$ 
near its starting point is $\approx 1/(\rho \overline{x(t)}) \sim 4 \delta^2 t^{-2 \delta T}/\rho$, a decay which exhibits some similarity with (\ref{survbiaslarget}) taken for $\rho \ll \delta^2$. In the other limit $\rho \gg \delta$ the effect of the bias is weak and one recovers the result  (\ref{survlarget}) for the unbiased case. 

\section{Diffusion front}
\label{diffusion}
\label{sec4}

Let us now compute the fully averaged diffusion front $\overline{P(x ,t|x_0, 0)}$, i.e. the probability that the particle has survived up to time $t$ and has moved from $x_0$ to $x$ in presence of the absorbers.

One first defines $\overline{P_L(x ,t|x_0 ,0)}$ in a fixed segment of size $L$ for the problem without absorbers, where here the average is only on $U(x)$ and one studies its Laplace transform $P(p,p_0,q) = \int_0^\infty dL \int_0^L dx_0 \int_0^L dx e^{-p x - p_0 x_0 - q L} \overline{P_L(x, t|x_0, 0)}$. Since in the effective dynamics the particle is at time $\Gamma=T \ln t$ at the bottom $x$ of the bond containing $x_0$, for each term containing $k\geq 1$ bulk valleys in the measure (\ref{fs}) one first defines:
\begin{eqnarray} \label{2n}
&& P^{2 n}_{2 k+2} = \langle \delta(x- \sum_{i=1}^{2n} \ell_i) \theta(\sum_{i=1}^{2 n-1} \ell_i \leq x_0 \leq x) \rangle_{L,k}  \\
&& P^{2 n+1}_{2 k+2} = \langle \delta(x- \sum_{i=1}^{2n} \ell_i) \theta(x \leq x_0 \leq \sum_{i=1}^{2 n+1} \ell_i )  \rangle_{L,k}
\label{2n+1}
\end{eqnarray}
corresponding to the initial condition $x_0$ being either on the $2n$-th bond or the $2n+1$-th, while $x$ is at the bottom of the $2n$-th bond, with $n=1,2,...k$ (see Fig. \ref{fig:1}). Note that the $P^{2n}$ corresponds to the particle moving to the right, i.e. $y=x-x_0>0$ while the $P^{2n+1}$ correspond to motion to the left $y<0$. Hence one separates $\overline{P_L(x, t|x_0, 0)} = 
\overline{P_L^+(x, t|x_0, 0)}+\overline{P_L^-(x, t|x_0, 0)}$ into the contribution of right motion ($P^+$) and left ($P^-$). 
In Laplace one has:
\begin{eqnarray}
\!\!\!\!\!\! \!\!\!\!\!\! P^{2 n}_{2 k+2}(p,p_0,q) &= & \frac{\bar \ell_\Gamma}{p_0} E^{-}_{p+p_0+q} (P^+_{p+p_0+q} P^-_{p+p_0+q})^{n-1} (P^+_{p+q} - P^+_{p+p_0+q}) P^-_q (P^+_q P^-_q)^{k-n} E^+_q \nonumber \\
\!\!\!\!\!\! \!\!\!\!\!\! P^{2 n+1}_{2 k+2}(p,p_0,q) & = &  \frac{\ell_\Gamma}{p_0} E^{-}_{p+p_0+q} (P^+_{p+p_0+q} P^-_{p+p_0+q})^{n-1} P^+_{p+p_0+q} (P^-_{q} - P^-_{p_0+q}) P^-_q (P^+_q P^-_q)^{k-n} E^+_q \nonumber 
\end{eqnarray}
Using the sum rule (\ref{normal}) and summing over $n-1$ and $k-n$ one finds:
\begin{eqnarray}
&& P^+(p,p_0,q) = \frac{1}{\bar \ell_\Gamma p_0 q (p+p_0+q)} \frac{(P^+_{p+q} - P^+_{p+p_0+q}) P^-_q}{E^+_{p+q+p_0} E^-_q} \\
&& P^-(p,p_0,q) = \frac{1}{\bar \ell_\Gamma p_0 q (p+p_0+q)} \frac{(P^-_{q} - P^-_{p_0+q}) P^+_{p+p_0+q}}{E^+_{p+q+p_0} E^-_q}
\end{eqnarray}
where $P^\pm(p,p_0,q)$ are the Laplace transforms of $\overline{P^\pm_L(x,  t|x_0 0)}$, corresponding to motion to the right $y>0$ and to the left $y<0$, respectively. 

We can now obtain the fully averaged diffusion front in our original model in the form:
\begin{eqnarray}
&& \overline{P(x_0+y , t|x_0 0)} = \theta(y) \overline{P^+(x_0+y , t|x_0 0)} + \theta(-y) \overline{P^-(x_0+y , t|x_0 0)}
\end{eqnarray}
as the sum of contributions of the $P^{2n}$ and $P^{2n+1}$ probabilities, respectively. The first one is obtained by setting
$p_0=-p$, i.e. one has:
\begin{eqnarray}  \label{right}
&& \int_0^\infty dy e^{-p y} \overline{P^+(x_0+y,  t|x_0 0)} \\
&& =  \rho^2 \int_0^\infty dL L e^{-\rho L} L^{-1} \int_0^L dx_0 \int_{-x_0}^{L-x_0} dy e^{-p y} \overline{P^+_L(x_0+y,  t|x_0 0)} \\
&& = \rho^2 P^+(p,-p,\rho) \nonumber = \frac{1}{\bar \ell_\Gamma p} \frac{(P^+_{\rho} - P^+_{p+\rho}) P^-_\rho}{E^+_{\rho} E^-_\rho}
\end{eqnarray}
while the second one is:
\begin{eqnarray} \label{left}
&& \int_{-\infty}^0 dy e^{p y} \overline{P^-(x_0+y,  t|x_0 0)} \\
&& =  \rho^2 \int_0^\infty dL L e^{-\rho L} L^{-1} \int_0^L dx_0 \int_{-x_0}^{L-x_0} dy e^{p y} \overline{P^-_L(x_0+y,  t|x_0 0)} \\
&& = \rho^2 P^-(-p,p,\rho) \nonumber =  \frac{1}{\bar \ell_\Gamma p} \frac{(P^-_{\rho} - P^-_{p+\rho}) P^+_\rho}{E^+_{\rho} E^-_\rho}
\end{eqnarray}

\subsection{symmetric Brownian landscape}

Let us first analyze this result for the symmetric case:
\begin{eqnarray} \label{propsym}
&&  \int_0^\infty dy e^{-p y} \overline{P^+(x_0+y,  t|x_0 0)} = \frac{\rho}{p} 
\frac{\cosh(\Gamma \sqrt{\rho})}{\sinh^2(\Gamma \sqrt{\rho})} (\frac{1}{\cosh(\Gamma \sqrt{\rho})}
-\frac{1}{\cosh(\Gamma \sqrt{p+\rho})}) \\
&& = \frac{\Gamma \sqrt{\rho}}{\sinh(2 \Gamma \sqrt{\rho})} [ 1 - \frac{1 + \Gamma \sqrt{\rho} ( 2 \tanh(\Gamma \sqrt{\rho}) -
\coth(\Gamma \sqrt{\rho} ))}{4 \rho} p + O(p^2))
\end{eqnarray}
and the same result for $P^-$ with $e^{- p y} \to e^{p y}$. 
The first term in the small $p$ expansion is half of the survival probability obtained in (\ref{surv1}), as expected. The second gives the average displacement 
{\it conditioned to having survived} (i.e. averaged over the walkers which have survived) is:
\begin{eqnarray}
&& \overline{|y|} =  \frac{1 + \Gamma \sqrt{\rho} ( 2 \tanh(\Gamma \sqrt{\rho}) -
\coth(\Gamma \sqrt{\rho} ))}{4 \rho}  
\end{eqnarray}
with $\Gamma=T \ln t$, with the following asymptotics respectively and small and large time:
\begin{eqnarray}
&& =  \frac{5}{12} T^2 \ln^2 t - \frac{29}{180} \rho T^4 \ln^4 t  + O(\ln^6 t) \quad , \quad T \ln t \ll 1/\sqrt{\rho} \\
&& =  \frac{1}{4 \sqrt{\rho}} T \ln t  + \frac{1}{4 \rho} - \frac{3}{2} \frac{T \ln t}{\sqrt{\rho}}  t^{- 2 T \sqrt{\rho}} + .. \quad , \quad T \ln t \gg 1/\sqrt{\rho} \label{asympt}
\end{eqnarray}
The small time behaviour $ \overline{|y|} =  \frac{5}{12} T^2 \ln^2 t $ is the one obtained in the Sinai model in absence of absorbers \cite{sinaiRSRG}. Note that the distance traveled by the surviving walkers only grows at large time as $\sim \xi(t)$, with:
\begin{eqnarray} \label{xi} 
&& \xi(t) = T  \ln t/\sqrt{\rho}
\end{eqnarray}
as these walkers are those which remain closer to their initial position (within a region whose size is still growing with time). This new length scale $\xi(t)$ is discussed further below.

Performing the inverse Laplace transform from (\ref{propsym}), one obtains the fully averaged diffusion front:
\begin{eqnarray} \label{propsym2}
&&
\!\!\!\!\!\! \!\!\!\!\!\! \overline{P(x_0+y,  t|x_0 0)} = \frac{ \pi \rho \cosh(\Gamma \sqrt{\rho})}{\sinh^2(\Gamma \sqrt{\rho})}  \sum_{n=0}^\infty \frac{(-1)^n (2n+1)}{\rho \Gamma^2 + \pi^2 (n+\frac{1}{2})^2} \exp( -  \rho |y|  - (n+\frac{1}{2})^2 \frac{\pi^2 |y|}{\Gamma^2})
\end{eqnarray}
with $\Gamma=T \ln t$. It is independent of the starting point and normalized to the survival probability $\int_{-\infty}^{+\infty} dy \overline{P(x_0+y,  t|x_0 0)}= S(t)$. In the limit $T \ln t \ll 1/\sqrt{\rho}$ where the probability that the walker has met an absorber is still small, one recovers the Kesten diffusion front \cite{Kesten}, derived in \cite{sinaiRSRG} using the RSRG (the same as the limit $\rho=0$).

Using the Poisson formula:
\begin{eqnarray} \label{poisson}
&& \sum_{n=-\infty}^\infty f(n) = \sum_{m=-\infty}^\infty \int_{-\infty}^{+\infty} dx f(x) e^{2 i \pi m x}
\end{eqnarray}
This can be rewritten in the dual form as:
\begin{eqnarray} \label{dual}
&&
\!\!\!\!\!\! \!\!\!\!\!\! \overline{P(x_0+y,  t|x_0, 0)} = S(t) \frac{  \sqrt{\rho} \cosh^2(\Gamma \sqrt{\rho})}{\sinh(\Gamma \sqrt{\rho})}   \sum_{m=0}^\infty \frac{2 m + 1}{ \pi^{1/2}} (-1)^m  \int_{|y|}^\infty \frac{dz}{z^{3/2}} e^{-\rho z  - \Gamma^2 \frac{(2 m+1)^2}{4 z}} \nonumber 
\end{eqnarray}
At large time $\Gamma=T \ln t \gg 1/\sqrt{\rho}$, this is dominated by the term $m=0$, which yields:
\begin{eqnarray} \label{propsymfin}
&&
\!\!\!\!\!\! \!\!\!\!\!\! \overline{P(x_0+y,  t|x_0, 0)} = S(t) \frac{ \sqrt{\rho}}{2  \pi^{1/2} } e^{\Gamma \sqrt{\rho}}  \int_{|y|}^\infty \frac{dz}{z^{3/2}} e^{-\rho z  - \frac{\Gamma^2}{4 z}} \nonumber 
\end{eqnarray}
Now one can perform the change of variable $y = \Gamma \tilde y/\sqrt{\rho}$ and $z = \Gamma \tilde z/\sqrt{\rho}$ in (\ref{propsymfin}) and taking the large $\Gamma$ limit one easily sees that either the saddle point point at $\tilde z=1/2$ is in the integration domain, and the result is constant, or it is not, in which case it gives a subdominant contribution. Hence in the large time limit the diffusion front for the rescaled variable $\tilde y= y/\xi(t)$ (where $\xi(t)$ is the new length scale defined above in (\ref{xi})) takes a remarkably simple form: it becomes {\it a uniform distribution} in the interval $|\tilde y| \in [0, 1/2]$ and zero outside:
\begin{eqnarray} \label{propsymfin3}
&&
\!\!\!\!\!\! \!\!\!\!\!\! \overline{P(x_0+y,  t|x_0, 0)} \approx S(t) ~ \frac{\sqrt{\rho}}{T \ln t} ~~ \theta(\frac{1}{2} - \frac{|y| \sqrt{\rho} }{T \ln t}) \quad , \quad T \ln t \gg 1/\sqrt{\rho}
\end{eqnarray}
where $\theta(x)=1$ for $x>0$ and $\theta(x)=0$ for $x<0$ is the Heaviside step function. The first
moment of this distribution reproduces (\ref{asympt}) above. Note that each integral in (\ref{dual}) has a saddle point at $z=(2 m+1) \Gamma/(2 \sqrt{\rho})$, resulting also in a step function probability for $|y|$ up to the saddle point, but the higher values $m \geq 1$ contribute only as a subdominant relative factor $t^{- 2 m T \sqrt{\rho}} \sim S(t)^m$ to the total probability, as compared to the main contribution. 

Equivalent manipulations can be performed on the Laplace transform. The limit for large $\Gamma$ of the expression (\ref{propsym}) reads:
\begin{eqnarray} \label{propsymfin2}
&&
\!\!\!\!\!\! \!\!\!\!\!\! \int_0^\infty e^{-p y} \overline{P(x_0+y,  t|x_0, 0)} = S(t) \frac{\sqrt{\rho}}{p \Gamma} ( 1 - e^{\Gamma (\sqrt{\rho} - \sqrt{\rho + p})}) 
\nonumber 
\end{eqnarray}
where we recall that $LT_{y \to p} \int_y^\infty dz f(z) = \frac{1}{p} (f(p=0)-f(p))$. In this expression further approximating $\sqrt{\rho + p}= \sqrt{\rho} + \frac{p}{2 \sqrt{\rho}} + .. $ one recovers the Laplace transform of (\ref{propsymfin3}) on the positive side.

To summarize, inside the sharp region (survival region) $- \frac{1}{2} < \tilde y=y/\xi(t) < \frac{1}{2} $ the diffusion front is uniform and decays uniformly as $S(t)$. Outside this region, there is a faster decay, which, in the variable $\tilde y$ can be described as power law in time with a continuously varying exponent, i.e. $\overline{P(x_0+y,  t|x_0, 0)} \sim t^{-a}$ with a decay exponent:
\begin{eqnarray} \label{propsymfin2}
&& a = T \sqrt{\rho} ( \tilde y + \frac{1}{4 \tilde y}  + 1 ) > 2 T \sqrt{\rho}
\end{eqnarray}
for fixed $\tilde y>1/2$.

It is interesting to note that for {\it free diffusion}, i.e. no random potential $U(x)=0$, an analogous phenomena occurs, i.e. there is also a surviving zone $- \xi_0(t) < y < \xi_0(t)$ with $\xi_0(t) \sim t^{1/3}$. It is sharp but not as sharp as what is found here for the Sinai diffusion. The details are recalled in Appendix B, where the analogies are emphasized.

\subsection{biased landscape}

The formula for the biased landscape obtained by inserting (\ref{formstart}) into our general result (\ref{right},\ref{left}) are more cumbersome, so we will not display them here, but concentrate on the large time limit.

Let us first display the normalizations, i.e. the survival joint probabilities {\it that the walker has survived and moved to the right} $P^+$ (resp. and moved to the left, $P^-$):
\begin{eqnarray} 
&& \!\!\!\!\!\! \!\!\!\!\!\! S^\pm(t) = \int dy \overline{P^\pm(x_0+y,  t|x_0 0)}= \frac{\delta^2 \Gamma \pm \delta+\Gamma \rho \mp \delta \Gamma  \sqrt{\delta^2+\rho} \coth \left(\Gamma \sqrt{\delta^2+\rho}\right)}{\sqrt{\delta^2+\rho}
   \sinh \left(2 \Gamma \sqrt{\delta^2+\rho}\right) \pm \delta (1-\cosh \left(2 \Gamma
   \sqrt{\delta^2+\rho})\right)} \nonumber
\end{eqnarray}
 with $\Gamma=T \ln t$. This yields the asymptotic behaviours respectively at small and large time:
\begin{eqnarray} 
 && S^\pm(t) = \frac{1}{2} \pm \frac{1}{3} \delta T \ln t - \frac{1}{3} \rho T^2 \ln^2 t \mp \frac{2 \delta^3 + 17 \delta \rho}{45} T^3 \ln^3 t + O(\ln^4 t) \\
   && = 2 (T \ln t \sqrt{\rho + \delta^2} \pm \frac{\delta}{\sqrt{\delta^2+\rho} \mp \delta}) ~  t^{- 2 T  \sqrt{\rho+\delta^2}} + ... \label{spmlarget} 
\end{eqnarray}
These two probabilities correctly sum up to the total probablity $S(t)=S^+(t)+S^-(t)$ that the walker has survived (\ref{survbias}). The validity is again small $\rho \sim \delta^2$ and the crossover then occurs at $T \ln t \sim 1/\sqrt{\rho + \delta^2}$. In the case $\rho \ll \delta^2$ the formula (\ref{spmlarget}) is valid only for much larger times, as discussed in section \ref{survivalproba}. 

Remarkably, the large time behaviour (\ref{spmlarget}) shows that the dominant term is the same, i.e. surviving particles have the same probability to have moved to the right than to the left. Let us recall that for $\rho=0$ one finds:
\begin{eqnarray} 
&& S^+_{\rho=0}(t) = P_{right}=\frac{1}{2} + \frac{1}{2} \coth(\delta T \ln t) - \frac{\delta T \ln t}{2 \sinh^2(\delta T \ln t)} \\
&& S^-_{\rho=0}(t) = P_{left}=1-P_{right} 
\end{eqnarray}
Hence $P_{left} \approx (2 \delta T \ln t -1) t^{-2 T \delta}$ and at large time almost all particules go to the right. One notices  that for $0 < \rho \ll \delta^2$ one has $S_{-}(t) \approx P_{left}$ hence in that regime the particles which go to the left are assured to survive, while the one which go to the right are assured to die. More generally, the ratio $S^+(t)/S^+_{\rho=0}(t) \approx S^+(t) \sim t^{- 2 T  \sqrt{\rho+\delta^2}}$ at large time, while for left going particles it is
$S^-(t)/S^-_{\rho=0}(t) \approx t^{- 2 T  (\sqrt{\rho+\delta^2} - \delta)}$, i.e it has a slower decay. Thus survival is always better among the particules which go to the left, but these are a small minority of them as the drift pushes most of them to the right. At the end the number of survivors is the same in the two vastly unequal right and left-moving populations.

Examination of the large time limit of the Laplace transform of $P^\pm$ yields a result very similar to the symmetric case, i.e. formula (\ref{propsymfin2}) replacing $\rho$ by $\rho + \delta^2$ everywhere. There is a surviving scale which reads now $\xi = (T \ln t)/\sqrt{\rho + \delta^2}$. Expanding further in the scaling region of  fixed $p \Gamma$, i.e. fixed $\tilde y=y/\xi$, and inverting the Laplace transform one finds again:
\begin{eqnarray} \label{propsymfinbias3}
&&
\!\!\!\!\!\! \!\!\!\!\!\! \overline{P(x_0+y,  t|x_0 0)} \approx S(t) ~ \frac{\sqrt{\rho + \delta^2}}{T \ln t} ~~ \theta(\frac{1}{2}   - \frac{|y| \sqrt{\rho + \delta^2}}{T \ln t}) \quad , \quad T \ln t \gg 1/\sqrt{\rho + \delta^2}
\end{eqnarray}
i.e a fully symmetric distribution for the surviving particles, with the same step function shape. Again this is valid when $\rho \sim \delta^2$. A more complicated crossover occurs in the large time regime when $\rho \ll \delta$ as the diffusion front first takes the forward exponential form $\sim \theta(x) \exp(x/\overline{x(t)})/\overline{x(t)}$ as in the absence of absorbers and finally crosses over to (\ref{propsymfinbias3}) at a much larger time which diverges as $\rho \to 0$. We will not study this crossover in details here.


\section{Average return probability} \label{return}
\label{sec5}

A slightly simpler quantity is the Laplace transform of the return to the origin $P(p_0,q) = \int_0^\infty dL \int_0^L dx_0  e^{- p_0 x_0 - q L} \overline{P_L(x_0, t|x_0, 0)}$. It is easily seen that one just needs to sum:
\begin{eqnarray}
&& P^{2 n}_{2 k+2}(p_0,q) =  \bar \ell_\Gamma E^{-}_{p_0+q} (P^+_{p_0+q} P^-_{p_0+q})^{n-1} P^+_{p_0+q} P^-_q (P^+_q P^-_q)^{k-n} E^+_q 
\end{eqnarray}
leading to:
\begin{eqnarray}
&& P(p_0,q) = \frac{1}{\bar \ell_\Gamma q (p_0+q)} \frac{P^+_{p_0+q} P^-_{q}}{E^+_{q+p_0} E^-_q}
\end{eqnarray}

The fully averaged return probability in the original model is given by summing over all initial positions with uniform weight and then over all segment sizes giving:
\begin{eqnarray}
&&  \overline{P(x_0, t|x_0, 0)}  = \rho^2 \int_0^\infty dL L e^{-\rho L} L^{-1} \int_0^L dx_0 \overline{P_L(x_0, t|x_0, 0)} =
\frac{1}{\bar \ell_\Gamma E^+_{\rho} E^-_{\rho}} - \rho 
\end{eqnarray}
This gives in the general case:
\begin{eqnarray}
&&  \overline{P(x_0, t|x_0, 0)}  = \frac{\rho + \delta^2}{\sinh^2( \Gamma \sqrt{\rho + \delta^2})} =  \frac{\rho + \delta^2 }{\sinh^2( \sqrt{\rho+ \delta^2} T \ln t )} 
\end{eqnarray} 
and the large time asymptotics:
\begin{eqnarray} \label{returnlarget} 
&& \overline{P(x_0, t|x_0, 0)}  \sim 4 (\rho+\delta^2)  ~  t^{-2 T \sqrt{\rho + \delta^2}} \quad , \quad T \ln t \ll 1/\sqrt{\rho + \delta^2} 
\end{eqnarray}
In the case of a symmetric landscape (zero bias) this is in agreement with the results of TH \cite{HT2} (to compare one should set $T=1$ here and $g=1/2$ there) in the limit of small $\rho$ (to leading order in $\rho$). Here we obtain the averaged return probability in the biased case as well, which exhibits a very similar expression. In particular, contrarily to other quantities discussed above it exhibits no complicated crossover for $\rho \ll \delta^2$. Indeed one can directly take the limit $\rho \to 0$ in the large time asymptotics (\ref{returnlarget}) and recover the probability of return to the origin of the biased Sinai random walk which is also a power law \cite{oldresults,Kesten,sinaiRSRG}. 

\section{Relaxation rates and physical picture of the power law decay} 
\label{sec6}

The aim of this Section is to explore in more details the distribution of relaxation times and obtain a simple physical picture for the power law relaxation. 

\subsection{lowest energy level} 

The (left and right) eigenfunctions of the diffusion (i.e. Fokker-Planck) operator $\phi_{L,R}(x)$ are related to the eigenfunctions of the Schrodinger operator $\psi(x)$ as $\phi_R(x)=e^{-U(x)/2 T} \psi(x)$ and $\phi_L(x)=e^{U(x)/2 T} \psi(x)$. This relation is compatible with the vanishing of both eigenfunctions at the edges of the segment $[0,L]$, as required here for perfect absorbers (Dirichlet boundary conditions). The eigenenergies $E_i$, i.e. the relaxation rates, are thus the same for both operators. 

The RSRG method allows to find all lowest eigenenergies as $\Gamma_i = - T \ln E_i$ where the $\Gamma_i$ are the scale of the successive decimations \cite{energyRSRG}. By convention we number $0<E_0<E_1,..$ the energies in increasing order, hence $\Gamma_0 > \Gamma_1 >...$ the corresponding decimation scales. Other methods to compute the statistics of the eigenenergies are presented in Ref. \cite{distributions} and in Ref. \cite{texierdos}, the latter being discussed at the end. 

We now recall the calculation of the probability distribution of the lowest energy level $E_0$, i.e. the smallest decay rate in a given sample of size $L$. It is associated to the decimation of the last bulk valley and occurs in the block $k=1$ of Fig. \ref{fig:1} when $\Gamma$ reaches the size $\Gamma_0$ of the smallest barrier of the two bonds, bringing the block $k=1$ to $k=0$. 
Hence the probability $Prob_L(\Gamma_0 < \Gamma)$ that the last decimation $\Gamma_0=- T \ln E_0$ in the segment $[0,L]$ has already occured at scale $\Gamma$, i.e. that $\Gamma_0<\Gamma$, is the total weight of the $k=0$ term in the measure (\ref{fs}):
\begin{eqnarray} \label{probg0} 
&& \int_0^\infty dL e^{-p L} Prob_L(\Gamma_0 < \Gamma) =  \bar l_\Gamma E_\Gamma^-(p)E_\Gamma^+(p) 
= \frac{1}{p} (1- P_\Gamma^-(p) P_\Gamma^+(p)) 
\end{eqnarray}
using the normalization (\ref{normal}). From this, the probability density $P_L(\Gamma_0) d\Gamma_0$ for $\Gamma_0= - T \ln E_0$ is obtained as $P_L(\Gamma_0) =  \partial_{\Gamma} Prob_L(\Gamma_0 < \Gamma)|_{\Gamma=\Gamma_0}$, and one finds:
\begin{eqnarray} \label{probg02} 
&& \int_0^\infty dL  e^{-p L} P_L(\Gamma_0) = \partial_{\Gamma_0}  \big( \frac{1}{(p + \delta^2) \coth^2(\Gamma_0 \sqrt{p+\delta^2}) - \delta^2} \big) 
\end{eqnarray}
Note that this formula was already obtained in \cite{sinaiRSRG} (Section VII-B) for the equilibration time, i.e the (inverse of the) energy of the first excited state, in a finite size sample with {\it reflecting} boundary conditions. However, in the limit studied here of a large sample and large times this turns out to be the same distribution (in a rescaled sense) as the lowest decay rate for the present problem, i.e. with {\it absorbing} boundaries \cite{footnote3}

In the absence of bias one finds:
\begin{eqnarray}
&& \int_0^\infty dL  e^{-p L} P_L(\Gamma_0) 
= - \partial_{\Gamma_0} \big( \frac{1}{p \cosh^2(\Gamma_0 \sqrt{p})} \big) =
- \frac{2}{\Gamma_0} \partial_p \big( \frac{1}{\cosh^2(\Gamma_0 \sqrt{p})} \big)
\end{eqnarray}
One can check that the relation $\frac{1}{\cosh^2(\Gamma \sqrt{p})} = E_\Gamma(p) + 2 p \partial_p E_\Gamma(p)$ holds in that case, and one finds, upon inverse Laplace transforming and using (\ref{direct}):
\begin{eqnarray} \label{e0proba} 
&& P_L(\Gamma_0) = \frac{2}{\Gamma_0} \big( L E_{\Gamma_0}(L) - 2 L \partial_L ( L E_{\Gamma_0}(L) ) \big) \\
&& = \frac{2 L}{\Gamma_0^3} \sum_{n=-\infty}^{+\infty} (2 \pi^2 (n+ \frac{1}{2})^2 \frac{L}{\Gamma_0^2} - 1) 
e^{- \pi^2 (n+ \frac{1}{2})^2 \frac{L}{\Gamma_0^2}} \\
&& =  \frac{8}{ L^{1/2} \sqrt{\pi}} \sum_{m=1}^{+\infty} (-1)^{m+1} m^2  
e^{- m^2 \frac{\Gamma_0^2}{L}}
\end{eqnarray}
using Poisson formula. 

Let us now recall that in a given segment $[0,L]$ the return probability is a trace over all eigenstates $\int_0^L dx_0 P_L(x_0,t|x_0 0) = \sum_n e^{-E^L_n t}$. Going back to the model with absorbers one averages over all segment sizes and starting points and one finds the relation:
\begin{eqnarray} \label{exact}
&&  \!\!\!\!\!\! \!\!\!\!\!\!  \overline{P(x_0, t|x_0, 0)} = \rho^2 \int_0^\infty dL L e^{-\rho L}  \int_0^L \frac{dx_0}{L} \overline{P_L(x_0,t|x_0 0)} = \rho^2 \int_0^\infty dL  e^{-\rho L}  \sum_n \overline{e^{-E^L_n t}} 
\end{eqnarray}
This quantity was computed in Section \ref{return}. To get physical insight into what controls the large time power law decay we will keep only the ground state in the sum, hence we get only a lower bound on the total decay. However we want to estimate the thermal average more precisely, to test the accuracy of the RSRG (i.e. to make sure we have not overlooked a process leading to a slower decay). From (\ref{exact}) we thus obtain in the absence of bias $\delta=0$:
\begin{eqnarray}
&& \!\!\!\!\!\! \!\!\!\!\!\!  \!\!\!\!\!\! \!\!\!\!\!\!  \overline{P(x_0, t|x_0, 0)} \geq
  \rho^2 \int_0^\infty dL e^{-\rho L} \int_0^\infty d\Gamma_0 e^{- t e^{-\Gamma_0/T}}  P_L(\Gamma_0) = \rho \int_0^\infty d\Gamma_0 e^{- t e^{-\Gamma_0/T}}  \partial_{\Gamma_0} \tanh^2(\Gamma_0 \sqrt{\rho}) \nonumber \\
&& =4 \rho  \int_0^1 dy  \frac{(1-y)}{(1+y)^3} e^{- t y^{1/(2 T \sqrt{\rho})} } \approx 4 \rho \Gamma(1+2 T \sqrt{\rho}) t^{-2 T \sqrt{\rho}} \label{as1}
\end{eqnarray}
setting $y = e^{-2 \Gamma_0 \sqrt{\rho}}$ and the last approximation is at large time. It is interesting to note that this is exactly the same prefactor for the large time decay as obtained by TH \cite{HT2}. The effective dynamics in the RSRG instead replaces $e^{- e^{x}} \to \theta(-x)$ function which then leads to:
\begin{eqnarray}
&& \overline{P(x_0 t|x_0 0)} \geq  \rho \int_{T \ln t}^\infty d\Gamma_0  \partial_{\Gamma_0} \tanh^2(\Gamma_0 \sqrt{\rho})  = \frac{\rho}{\cosh^2(T \ln t \sqrt{\rho}) } 
 \approx 4 \rho t^{-2 T \sqrt{\rho}} \label{as2} 
\end{eqnarray}
and the last approximation is at large time. As discussed above the RSRG is only expected to be accurate for large $L_\rho$ i.e. small $\rho$, in which case the two asymptotic behaviours (\ref{as1}) and (\ref{as2}) cannot be distinguished. Taken together the results  (\ref{as1}) and (\ref{as2}) compared to (\ref{returnlarget}) (in the absence of bias) show that (i) the asymptotic time decay can be correctly obtained by retaining only the longest relaxation time in each segment between absorbers and then averaging over the segments (ii) that the particles which jump anomalously late are sufficiently rare not to affect the leading asymptotics at small $\rho$ \cite{footnote2}

\subsection{saddle point picture, universality and rare events} 

We can now clarify the physical origin of the power law decay. First we can concentrate on the longest relaxation time (i.e. smallest energy) in each segment between absorbers $t_0 = e^{\Gamma_0/T}$. We must explain how a power law distribution of relaxation times of the form $\sim dt_0/t_0^{1+2 T \sqrt{\rho}}$ is generated by the combination of absorbers and disorder. We note from (\ref{e0proba}) that the distribution of barriers $\Gamma_0$ decays as $\sim \exp(- \Gamma_0^2/L)$ for large barriers, i.e. the distribution of relaxation time at fixed $L$ has a log-normal tail. Averaging over the various sizes $L$ of the segments between absorbers indeed produces the power law spectrum:
\begin{eqnarray} \label{saddlep}
&& \int dL e^{- \rho L } e^{-  \Gamma_0^2/L } \sim e^{- 2 \sqrt{\rho} \Gamma_0} \sim t_0^{-2 T \sqrt{\rho}} 
\end{eqnarray}
via a saddle point at $L=\xi(t_0)=\Gamma_0/\sqrt{\rho} = (T \ln t_0)/\sqrt{\rho}$ which is the optimal size of segments contributing to relaxation times $t_0$. Note that this is also the survival length scale associated to time $t_0$ 
already discussed above, and four times the one obtained in (\ref{asympt}), which is quite consistent.

These considerations also allow to discuss the universality of the present results, beyond the Brownian energy landscape. In problems of large time decay in random media involving some saddle point argument, one is always at risk of neglecting a slower process due to very rare events. By using here the RSRG fixed point based on the Brownian landscape we assume somehow that the saddle point indeed occurs in the scaling region $\Gamma_0 \sim \sqrt{L}$ where the probability tail $e^{-  \Gamma_0^2/L }$ is valid (i.e. large but still typical barriers). In fact, from the saddle point in (\ref{saddlep}) we need more, we need that this tail holds where the saddle point occurs, i.e. at $\Gamma_0 \sim \sqrt{\rho} L$ (barriers larger than typical). Hence it is safe to expect that it holds only for small $\rho$. To test the validity for larger $\rho$, one may consider e.g. the most extreme events where the landscape $U(x)$ between two scatterers has an unusually low minimum with a barrier to exit of order $\sim -L/2$. One can evaluate the probability of such an event to $\sim e^{- L \ln 2}$ (e.g. for a $\pm 1$ discrete random walk $U(x)$), and the total contribution of all such events as $\sim \int dL \exp( - (\rho + \ln 2) L - t/e^{L/2 T} ) \sim t^{- 2 T (\rho + \ln 2)}$. This is again a power law, which turns out to be faster than the power law (\ref{saddlep}) for any $\rho$. Hence the Brownian landscape is stable at least
to such events for a discrete bare model. Generally these are events where barriers are smaller than expected from a Brownian landscape (e.g. there is a maximum possible barrier $-L_0/2$ in a sample of size $L_0$). The opposite case, where there can be bigger barriers than for the Brownian landscape, may be problematic. For instance, although the RSRG fixed point is also valid for power law distributions of barriers in the original model with fast enough decaying tails \cite{fisher1}, it is not clear whether, in that case, these tails may not always lead to a slower decay. We leave this question to future investigation.

Another extension is the case where $p(L)$, the distribution of segment sizes, is not simply an exponential. Although exact calculations can be performed using the above methods, we can sketch the result of the saddle point argument. From (\ref{saddlep}) one sees that if $p(L)$ is delta-peaked on a given $L_0$ the large time decay will be as $\sim \exp( - (T^2 \ln^2 t)/L_0)$. But then, for a discrete model, it must change at very large time, i.e. $T \ln t \sim L_0$, to a faster $e^{- L_0 \ln 2 - t/e^{L_0/2T}}$ decay as the distribution of relaxation times hits its large time cutoff (i.e. the largest possible barrier in a sample of size $L_0$). If now the tail is  $p(L) \sim \exp(-(\rho L)^\gamma)$ with $\gamma>1$ one expects a faster decay than in the exponential case, i.e. as $\sim \exp(- c_\gamma (\sqrt{\rho} T \ln t)^{2 \gamma/(\gamma+1)})$ via a similar saddle point argument, and $c_\gamma=(1+\gamma) \gamma^{-\gamma/(\gamma+1)}$, which reproduces the previous result for the Poisson case $\gamma=1$. Again one must check a posteriori that the Brownian landscape tail $e^{-  \Gamma_0^2/L }$ holds for the given model around the saddle point $L \sim \Gamma_0^{2/(1+\gamma)}$. And one finds, as before, that for a discrete model, one can hope that the saddle point behavior holds only up to $T \ln t \sim (1/\rho)^{\gamma/(\gamma-1)}$ when the maximal barrier is reached. For the Poisson case $\gamma=1$, the saddle point picture nicely holds for all large times. Finally the above stretched-log decay formula should also holds for $\gamma<1$ when the decay becomes slower than a power law. 

\subsection{discussion of the energy spectrum} 

Before we close this section, let us make a few comments. First, one can also compute from the RSRG the distribution of the $k$-th eigenvalue $E_k$, from the weight of the $k$-th block in the measure (\ref{fs}):
\begin{eqnarray}
&& \int_0^\infty dL e^{-p L} Prob_L(\Gamma_k < \Gamma) =  \bar l_\Gamma E_\Gamma^-(p)E_\Gamma^+(p) \sum_{j=0}^k (P_\Gamma^-(p)P_\Gamma^+(p))^j \\
&& = \frac{1}{p} (1 - (P_\Gamma^-(p)P_\Gamma^+(p))^{k+1})
\end{eqnarray}
In the symmetric case using $\Gamma \partial_\Gamma = 2 p \partial_p$ acting on the fixed point probability, this gives:
\begin{eqnarray} \label{gen}
&& P_L(\Gamma_k) = \frac{2}{\Gamma_k} L ~P_{\Gamma_k}^{*2k+2}(L) 
\end{eqnarray}
i.e. $L$ times the $2k+2$-th convolution of a bulk bond length distribution, i.e. the distribution of size of a renormalized valley with $2k+2$ bonds. 

Second, let us comment on a relation between the method used by Texier in Ref. \cite{texierdos} and the RSRG. As discussed in \cite{energyRSRG} for the infinite sample, i.e. the bulk, the zeroes of the eigenstates with energies $T \ln E_i < \Gamma$ are on hill tops of the renormalized landscape at scale $\Gamma$,  hence the distance between the successive zeroes are given by the length of the valley $\ell_1+\ell_2$. This is precisely the result of Ref. \cite{texierdos}: the $P(\ell)$ there (formula (104) and (154) there) is the distribution of length of a bulk valley i.e. the convolution of two $P(\ell)$ here (while here $P(\ell)$ denote the distribution of length of a bond, also called $\Lambda$ in Ref.\cite{texierdos}). This was obtained there from the statistics of the escape time of the stochastic Ricatti equation, while in the RSRG is it obtained as the fixed point of an asymptotically exact decimation method. Hence as far as bulk properties are concerned there seems to be perfect agreement between the two methods. The formula (\ref{e0proba}) coincides with the result of Ref. \cite{texierdos} (formula (123) and (110) there). 

These arguments must be modified to take into account the edge of the system. In the RSRG this is described by the edge bonds. The arguments of \cite{texierdos} appear also to be correct for Dirichlet boundary conditions as required here. This is because the length of the system $L$ was chosen there to be the interval between two zeroes of the eigenfunctions. As a result one can indeed check that the formula (\ref{e0proba}) coincides with the result of Ref. \cite{texierdos} (formula (123) and (110) there). More generally formula (\ref{gen}) for the distribution of $E_k$ coincides with formula (132) in Ref. \cite{texierdos}. Given this agreement between the two methods for calculation of the distribution of the eigenenergies with Dirichlet boundary conditions, it is quite natural that we find agreement for all results which can be compared.

\section{Meeting time of two particles}
\label{meeting}
\label{sec7}

The RSRG allows to compute a host of more exotic properties of the trajectories, such as aging, return to the origin, persistence etc.. Here we will restrict to consider the meeting properties of two particles. 

Consider now two particles distant by $y_0>0$ at time $t=0$, and let us compute the probability $M(y_0,t)$ that
at time $\Gamma= T \ln t$ (i) they are both alive AND (ii) they have already met at least once. This quantity is simpler to compute since it requires the two particles to have started in the same segment, denoted $[0,L]$ between two adjacent absorbers (otherwise they never meet). Note, en passant, that the probability that two particles start in the same interval of size $L$ is $\rho^2 e^{- \rho L} (1-\frac{y_0}{L}) \theta(L<y_0)$, hence a total probability $e^{- \rho y_0}$ to start in the same interval. Even if they start in the same interval, they may still be absorbed before meeting.

As discussed in \cite{sinaiRSRG} having met is equivalent to the initial positions of the particle, let us call them $x_0$ and $x_0+y_0$ belonging to the same renormalized valley at $\Gamma$. The probability $M(y_0,t)$ can thus be obtained from the probability $M_L(y_0,t)$ that, in a sample of size $L$, the interval $[x_0,x_0+y_0]$ is contained in a single renormalized bulk valley (denoted $2n,2n+1$ below). One has:
\begin{eqnarray}
M_L(y_0,t) = \sum_{k=1}^\infty \sum_{n=1}^{n=k} \int_0^L \frac{d x_0}{L} \langle \theta(\sum_{i=1}^{2 n-1} \ell_i \leq x_0) 
\theta(x_0+y_0 \leq \sum_{i=1}^{2 n+1} \ell_i) \rangle_{L,k}
\end{eqnarray}
in the notations of Section \ref{diffusion}, given that the measure on $x_0$ is uniform in the interval. As usual one computes:
\begin{eqnarray}
&& M(q,p,t)= \int_0^\infty dL L e^{-q L} \int_0^\infty e^{-p y_0} M_L(y_0,t) 
\end{eqnarray}
Inserting the finite size measure (\ref{fs}), using that $L=\sum_{i=1}^{2 k+2} \ell_i$, performing the integration over $x_0$ and $y_0$, summing over $n-1$ and $k-n$ and using the normalization identity, one finds, very much as in Section \ref{diffusion}:
\begin{eqnarray}
&&  \!\!\!\!\!\! \!\!\!\!\!\!  M(q,p,t)= \frac{1}{p^2} ( N(p) - N(0) - p N'(0) ) \quad , \quad  N(p) =  \frac{P_\Gamma^+(p+q) P_\Gamma^-(p+q)}{\bar \ell_\Gamma q^2 E_\Gamma^+(q) E_\Gamma^-(q)}
\end{eqnarray}
Upon averaging over all segment sizes one finds that $\rho^2 M(\rho,p,t)$ is the Laplace transform of $M(y_0,t)$ and we obtain our final result for the model with absorbers:
\begin{eqnarray}
&&  \!\!\!\!\!\! \!\!\!\!\!\!  M(y_0,t) = LT^{-1}_{p \to y_0}  \frac{1}{p^2} ( f(p) - p f'(0) ) \quad , \quad   f(p) =  \rho - \frac{ (p+\rho)  E_\Gamma^+(p+\rho) E_\Gamma^-(p+\rho)}{ E_\Gamma^+(\rho) E_\Gamma^-(\rho)} 
\end{eqnarray} 
with $\Gamma=T \ln t$ and $- f'(0)=S(t)$ is the survival probability computed in Section \ref{survivalproba}. One finds that in the limit $\rho=0$ one recovers the result of \cite{sinaiRSRG}. 

In the symmetric case one finds:
\begin{eqnarray} \label{meeting1}
&& M(y_0,t) = \frac{2 \Gamma \sqrt{\rho}}{\sinh(2 \Gamma \sqrt{\rho})} - \frac{\rho}{\sinh^2(\Gamma \sqrt{\rho})} y_0 + \frac{\rho}{\tanh^2(\Gamma \sqrt{\rho})} 
\int_0^{y_0} dy \int_0^y dz  H(z)   \nonumber \\
&&  \!\!\!\!\!\! \!\!\!\!\!\!  \!\!\!\!\!\! \!\!\!\!\!\! = \frac{8 \rho}{ \tanh^2(\Gamma \sqrt{\rho})}  \sum_{m=-\infty}^{+\infty}  \frac{\pi^4 (2 m+1)^4 y_0 + 2 \Gamma^2 \pi^2 (2 m+1)^2 (3 + 2 \rho y_0) - 8 \Gamma^4 \rho}{(4 \Gamma^2 \rho + \pi^2 (2 m+1)^2)^3} e^{- (\frac{\pi^2}{\Gamma^2} (m+\frac{1}{2})^2 + \rho) y_0 } \nonumber
\end{eqnarray} 
with:
\begin{eqnarray} \label{hz}
&& \!\!\!\!\!\! H(z) = LT^{-1}_{p \to z} \frac{1}{\cosh^2(\Gamma \sqrt{p+\rho})} = \frac{1}{2 \Gamma^2} \sum_{m=-\infty}^{+\infty} 
(\pi^2 (2 m+1)^2 \frac{z}{\Gamma^2} - 2) e^{- (\frac{\pi^2}{\Gamma^2} (m+\frac{1}{2})^2 + \rho) z } \nonumber \\
&& = \frac{4 \Gamma}{\sqrt{\pi} z^{3/2}} \sum_{n=1}^\infty (-1)^{n+1} n^2 e^{- n^2 \frac{\Gamma^2}{z} - \rho z}
\end{eqnarray} 
One can check that formula (\ref{meeting}) for $M(y_0,t)$ matches the result (126) for the meeting probability (denoted there $1-F_{y_0}(\Gamma)$) in  \cite{sinaiRSRG} in the limit $\rho=0$ (equivalently the leading order in the limit $\rho \ll y_0 \sim \Gamma^2$). For small times $\Gamma^2 \ll y_0 \sim 1/\rho$ one finds:
\begin{eqnarray}
&& M(y_0,t) \approx \frac{16}{\Gamma^2} 
\frac{8 y_0}{\pi^2} e^{- \pi^2 y_0/4 \Gamma^2}  e^{- \rho y_0 } 
\end{eqnarray} 
with $\Gamma=T \ln t$. One may interpret the extra factor due to the absorbers as the probability, discussed above, that the two particles belong to the same interval between absorbers: if they are, then the finite size of this interval appears to play only a sub-leading role at short times.

Let us now study the large time limit defined as $\Gamma^2 \gg 1/\rho$ and $\Gamma^2 \gg y_0$ with $\Gamma=T \ln t$. Then we can retain
only the term $n=1$ in (\ref{hz}) and use again the survival length scale $\xi(t)=T \ln t/\sqrt{\rho}$ introduced above in (\ref{xi}) defining $\tilde y_0 = y_0/\xi(t)$. Then we obtain:
\begin{eqnarray} \label{meetingt}
&& M(y_0,t) \approx 4 \Gamma \sqrt{\rho}  (1 - \tilde y_0) e^{-2 \Gamma \sqrt{\rho}} + 4 \rho \Gamma
(\frac{\Gamma}{\sqrt{\rho}})^{1/2} \int_0^{\tilde y_0} d \tilde y \int_0^{\tilde y}  \frac{d \tilde z}{\sqrt{\pi} \tilde z^{3/2}}  e^{- \Gamma \sqrt{\rho} (\frac{1}{\tilde z} +  \tilde z) }  
\end{eqnarray}
For large $\Gamma \sqrt{\rho}$ the last term has a saddle point at $\tilde z=1$. It is subdominant for  $0<y_0<T \ln t/\sqrt{\rho}$ and in that case the large time decay is:
\begin{eqnarray}
&& M(y_0,t) \approx 4 T \ln t \sqrt{\rho} ( 1 - \tilde y_0) t^{-2 T \sqrt{\rho}} = (1-\tilde y_0) S(t)
\end{eqnarray}
Hence only the amplitude is changed with respect to the (one particle) survival probability. If the two particle are further apart, i.e. for $y_0>T \ln t/\sqrt{\rho}$ the leading saddle point contribution cancels exactly the first two terms in (\ref{meetingt}). What remains is thus a faster decay. To study it let us start from the exact formula for the derivative (valid in all time regimes), easily obtained from (\ref{hz}):
\begin{eqnarray} \label{meeting1}
&& \partial_{y_0} M(y_0,t) = -  \frac{\rho}{\tanh^2(\Gamma \sqrt{\rho})} \int_{y_0}^\infty dz  H(z)  
\end{eqnarray} 
For large $\Gamma \sqrt{\rho}$ one has thus:
\begin{eqnarray}
&& \partial_{\tilde y_0} M(y_0,t) \approx - 4 \rho \Gamma
(\frac{\Gamma}{\sqrt{\rho}})^{1/2} \int^\infty_{\tilde y_0} \frac{d \tilde z}{\sqrt{\pi} \tilde z^{3/2}}  e^{- \Gamma \sqrt{\rho} (\frac{1}{\tilde z} +  \tilde z) }  
\end{eqnarray}
hence for $\tilde y_0 >1$ we find a decay $M(y_0,t) \sim t^{- T \sqrt{\rho} (\tilde y_0 + \frac{1}{\tilde y_0})}$ with an exponent which increases continuously with the distance. The large time behaviour is thus similar to the one found for the diffusion front in Section \ref{diffusion}. It can also be understood from a related calculation of the survival probability of two particles, performed in the Appendix A. 

\section{Distance traveled before absorption}
\label{sec8}

Here we compute the distribution $Q(z)$ of the total distance traveled, $z$, by a walker before
it is absorbed. This quantity was obtained for free diffusion (i.e. in the absence of a random potential)
in Ref. \cite{sire1}. We consider a uniform density for the initial position of the walker. 

Consider a segment $[0,L]$ between two successive impurities. To each end of the segment corresponds an absorbing zone of length $\ell_1$, $\ell_2$ respectively (see Fig. \ref{fig:1}), which determines if the walker is absorbed to the left or to the right, respectively. For a fixed $L$ the probability distribution of $(\ell_1,\ell_2)$ is:
\begin{eqnarray}
&& q_L(\ell_1,\ell_2) = \lim_{\Gamma \to \infty} \bar \ell_\Gamma E_\Gamma^{-}(\ell_1) E_\Gamma^{+}(\ell_2) \delta(\ell_1+\ell_2-L) = e^-(\ell_1) e^+(\ell_2)  \delta(\ell_1+\ell_2-L) 
\end{eqnarray}
with, in Laplace and in real space, respectively:
\begin{eqnarray}
&& e^\pm(p) = \frac{1}{\sqrt{p+\delta^2} \mp \delta} \quad , \quad e^\pm(\ell) = \frac{e^{-\delta^2 \ell}}{\sqrt{\pi \ell}} - \delta ~ \phi(\delta \sqrt{\ell}) + \delta \pm \delta \\
&& \phi(x) = \int_x^\infty dy \frac{2}{\sqrt{\pi}} e^{-y^2} \label{error} 
\end{eqnarray}
$\phi(x)$ being the complementary error function, and $e^\pm(\ell)=1/\sqrt{\pi \ell}$ in  the symmetric case $\delta=0$. Not surprisingly, the above probability can be interpreted in terms of extremal statistics of the (landscape) Brownian motion \cite{sinaiRSRG,greginprep}. It is useful to define the full Laplace transform:
\begin{eqnarray}
&&  \!\!\!\!\!\! \!\!\!\!\!\! \!\!\!\!\!\! \!\!\!\!\!\! q(p,p_1,p_2) =  \int_0^\infty dL e^{-p L} \int_0^\infty d\ell_1 \int_0^\infty d\ell_2 e^{-p_1 \ell_1 - p_2 \ell_2} q_L(\ell_1,\ell_2) = e^{-}(p+p_1) e^{+}(p+p_2)
\end{eqnarray}
which is correctly normalized to $q(p,0,0)=1/p$, i.e. $\int_{\ell_1,\ell_2} q_L(\ell_1,\ell_2) = 1$ for each $L$. From this quantity we can obtain a number of observables for our model:

First we obtain the probability that an interval between successive absorbers (chosen at random among all such intervals with equal probability, i.e. length distribution $\rho e^{-\rho L}$) has absorbing zones $(\ell_1,\ell_2)$:
\begin{eqnarray}
&& \tilde q_0(\ell_1,\ell_2) = LT^{-1}_{p_1 \to \ell_1,p_2 \to \ell_2} \rho q(\rho,p_1,p_2) = \rho e^{-\rho(\ell_1+\ell_2)} e^{-}(\ell_1) e^{+}(\ell_2)
\end{eqnarray}

Second, we obtain the probability that a walker chosen at random (i.e. with uniform density of initial position) belongs to an 
interval (i.e. of length distribution $\rho^2 L e^{-\rho L}$) with absorbing zones $(\ell_1,\ell_2)$:
\begin{eqnarray}
&&  \!\!\!\!\!\! \!\!\!\!\!\! \!\!\!\!\!\! \!\!\!\!\!\!  \tilde q_1(\ell_1,\ell_2) = LT^{-1}_{p_1 \to \ell_1,p_2 \to \ell_2} \rho^2 (\partial_{p_1} +  \partial_{p_2}) q(\rho,p_1,p_2)
= \rho^2 (\ell_1 + \ell_2) e^{-\rho(\ell_1+\ell_2)} e^{-}(\ell_1) e^{+}(\ell_2)
\end{eqnarray}

Finally let us denote $z$, the distance, counted with its sign, traveled up to absorption. One has $z=- x_0$ if $x_0 \in [0,\ell_1]$ (right absorption) and $z=\ell_1+\ell_2-x_0=L-x_0$ if
$x_0 \in [\ell_1,L]$ (left absorption). Hence one finds:
\begin{eqnarray}
&&  \!\!\!\!\!\! \!\!\!\!\!\! \!\!\!\!\!\! \!\!\!\!\!\!  Q(z) = \rho^2 \int_0^\infty dL L e^{-\rho L}  \int_0^\infty d\ell_1 \int_0^\infty d\ell_2  
( \frac{1}{L} \theta(-z) \theta(|z| < \ell_1) + \frac{1}{L} \theta(z) \theta(|z| < \ell_2)  ) q_L(\ell_1,\ell_2) \nonumber \\
&& = \theta(- z) Q^-(|z|) + \theta(z) Q^{+}(z) 
\end{eqnarray}
where we used that $\frac{\ell_1}{L} \int_0^{\ell_1} \frac{dx_0}{\ell_1} \delta(z+x_0) =  \frac{1}{L} \theta(-z) \theta(|z| < \ell_1)$
is the probability that $x_0$ belongs to the first bond and that $z=-x_0$. One thus has, in Laplace, $Q^-(p_1) = \rho^2 \frac{1}{p_1} (q(\rho,0,0) - q(\rho,p_1,0))$ and $Q^+(p_2) = \rho^2 \frac{1}{p_2} (q(\rho,0,0) - q(\rho,0,p_2)) $, i.e.:
\begin{eqnarray}
&& Q^\pm(p) =  \rho^2 \frac{1}{p} ( \frac{1}{\rho} -  \frac{1}{(\sqrt{\rho+\delta^2} \pm \delta) (\sqrt{\rho+p+\delta^2} \mp \delta)} )
\end{eqnarray}
The total probability of being absorbed to the left is $Q_L=Q^-(0)=\frac{1}{2} (1-\frac{\delta}{\sqrt{\rho+\delta^2}})$ and to the
right $Q_R = Q^+(0)=\frac{1}{2} (1+\frac{\delta}{\sqrt{\rho+\delta^2}})$, which should be seen as a scaling function of
$\delta/\rho$ in the limit where both are small. Inverse Laplace transform gives:
\begin{eqnarray}
&& Q^\pm(|z|) = \frac{\rho}{\sqrt{\rho+\delta^2} \pm \delta}  ( \delta e^{- \rho |z|} (1 \pm 1 -\phi(\delta \sqrt{|z|})) + \sqrt{\rho + \delta^2} \phi(\sqrt{\rho + \delta^2} \sqrt{|z|}) ) 
\end{eqnarray}
with $\phi(x)$ being the complementary error function given in (\ref{error}). In the symmetric case, $\delta=0$, the probability of the distance traveled $z$, counted with its sign is:
\begin{eqnarray}
&& Q(z) = \rho \phi( \sqrt{\rho |z|})
\end{eqnarray}
normalized on the real axis $z \in ]-\infty,+\infty[$. Since this quantity integrates over all the history of the walker, the only
relevant length scale is $L_\rho=1/\rho$. The distance traveled by walkers absorbed up to time $t$ can also be obtained by similar methods. 

\section{Schrodinger Green's function}
\label{sec9}

Let us first recall the idea of the calculation in the absence of absorbers. The Green's function of the
Schrodinger operator is (we set $T=1$ in this Section):
\begin{eqnarray}  \label{relation}
&& G_L(x,t|x_0,0) = e^{ (U(x)-U(x_0))/2} P_L(x,t|x_0,0)
\end{eqnarray}
as discussed above this extends to finite size $L$, consistent with the vanishing of the wavefunctions at the boundaries. The infinite sample case was discussed in \cite{sinaiRSRG} Section VIII. Because $P_L(x,t|x_0,0)$ behaves as $\exp(-U(x))$ near the actual position of the particle at time $t$, i.e. the end of a renormalized bond, it is clear that the non vanishing contributions to $G_L(x,t|x_0,0)$ will come only from (rare) bonds which have a second minimum degenerate with the one at the bond edge (and $x,x_0$ near each of them, the order not counting, as $G_L$ is a symmetric function of its arguments). Hence we introduced there the probability $R_\Gamma(\ell,z)$ that a renormalized bond at scale $\Gamma$ has length $\ell$ and a distinct degenerate minimum at distance $z>0$ from the absolute minimum in the bond. Of course $z$ must be of order $\Gamma^2$, the many quasi-degenerate minima around $z=0$ are counted as a different (a delta-function) contribution (see discussion of precise definitions in Appendix E of \cite{sinaiRSRG}. It was shown there that:
\begin{eqnarray}
&& R^\pm_\Gamma(\ell,z) = P^\pm_\Gamma(\ell-z) g(z) \quad , \quad g(z) = \frac{1}{\Gamma^3} G(z/\Gamma^2) e^{- z \delta^2} \\
&& G(X) = 4 \pi^2 \sum_{n=1}^\infty n^2 e^{- X n^2 \pi^2} = \frac{1}{\sqrt{\pi} X^{3/2}} \sum_{m=-\infty}^{+\infty} (1 + \frac{2 m^2}{X}) e^{-m^2/X} 
\end{eqnarray}
where $g(z)$ does not depend on the direction of the bias. 

We can now compute $\overline{G_L(x,t|x_0,0)}$ (the average is only over $U(x)$ in a finite segment $[0,L]$) by a simple modification of the calculation of $\overline{P_L(x,t|x_0,0)}$ in Section \ref{diffusion}. We simply need to replace the bulk bond in the renormalized landscape at $\Gamma=T \ln t$ containing the initial position $x_0$ by a degenerate bond, i.e. a $R$ bond (if $x_0$ belongs to on an edge bond the particle is absorbed, hence we never have to replace those). Hence in the average (\ref{2n}) we need to write now:
\begin{eqnarray} \label{2n}
&& \!\!\!\!\!\! \!\!\!\!\!\! \!\!\!\!\!\! \!\!\!\!\!\! G^{2 n}_{2 k+2} = \langle \delta(x- \sum_{i=1}^{2n} \ell_i) \delta(x-x_0-z) \rangle_{L,k}  \quad , \quad  G^{2 n+1}_{2 k+2} = \langle \delta(x- \sum_{i=1}^{2n} \ell_i) \delta(x_0-x-z) \rangle_{L,k}
\label{2n+1}
\end{eqnarray}
where $x$ is at the bottom of the $2n$-th bond and $x_0$ either on the $2n$-th bond (first line) or the $2n+1$ (second line). Hence the average is such that in the first case the $2n$-th bond is a $R$ bond while in the second case the bond $2n+1$ is a $R$ bond. Note that strictly speaking $x$ and $x_0$ can be interchanged (when a bond is degenerate there is an ambiguity on the definition of the two bonds forming the valley). We find it more convenient, and equivalent to symmetrize at the end.

Introducing now
$G(p,p_0,q) = \int_0^\infty dL \int_0^L dx_0 \int_0^L dx e^{-p x - p_0 x_0 - q L} \overline{G_L(x t|x_0 0)}$ and using that in Laplace $R^\pm_{p,q}=P^\pm_{p} g_{q+p}$, we obtain
\begin{eqnarray}
\!\!\!\!\!\! \!\!\!\!\!\! G^{2 n}_{2 k+2}(p,p_0,q) &= & \bar \ell_\Gamma E^{-}_{p+p_0+q} (P^+_{p+p_0+q} P^-_{p+p_0+q})^{n-1} P^+_{p+p_0+q} g_{p+q} P^-_q (P^+_q P^-_q)^{k-n} E^+_q \nonumber \\
\!\!\!\!\!\! \!\!\!\!\!\! G^{2 n+1}_{2 k+2}(p,p_0,q) & = & \ell_\Gamma  E^{-}_{p+p_0+q} (P^+_{p+p_0+q} P^-_{p+p_0+q})^{n-1} P^+_{p+p_0+q} P^-_{q} g_{p_0+q}  (P^+_q P^-_q)^{k-n} E^+_q \nonumber 
\end{eqnarray}

Using the sum rule (\ref{normal}) and summing over $n-1$ and $k-n$ one finds:
\begin{eqnarray}
&& G^+(p,p_0,q) = \frac{1}{\bar \ell_\Gamma q (p+p_0+q)} \frac{P^+_{p+p_0+q} g_{p+q} P^-_q}{E^+_{p+q+p_0} E^-_q} \\
&& G^-(p,p_0,q) = \frac{1}{\bar \ell_\Gamma q (p+p_0+q)} \frac{P^+_{p+p_0+q} g_{p_0+q} P^-_{q} }{E^+_{p+q+p_0} E^-_q}
\end{eqnarray}
where $G^\pm(p,p_0,q)$ are the Laplace transforms of $\overline{G^\pm_L(x,  t|x_0 0)}$, corresponding to motion to the right $y=x-x_0>0$ and to the left $y<0$, respectively (given that these will be later symmetrized). They precisely correspond to the exchange of $p$ and $p_0$. We can now obtain the fully averaged diffusion front in our original model in the form:
\begin{eqnarray}
&& \overline{G(x_0+y , t|x_0 0)} = \theta(y) \overline{G^+(x_0+y , t|x_0 0)} + \theta(-y) \overline{G^-(x_0+y , t|x_0 0)}
\end{eqnarray}
as the sum of contributions of the $G^{2n}$ and $G^{2n+1}$ probabilities, respectively. The first one is obtained by setting
$p_0=-p$, i.e. one has:
\begin{eqnarray}  
&& \int_0^\infty dy e^{-p y} \overline{G^+(x_0+y,  t|x_0 0)} \\
&& =  \rho^2 \int_0^\infty dL L e^{-\rho L} L^{-1} \int_0^L dx_0 \int_{-x_0}^{L-x_0} dy e^{-p y} \overline{G^+_L(x_0+y,  t|x_0 0)} \\
&& = \rho^2 G^+(p,-p,\rho) \nonumber 
= \frac{1}{\bar \ell_\Gamma} \frac{P^+_{\rho} g_{p+\rho} P^-_\rho}{E^+_{\rho} E^-_{\rho}} 
\end{eqnarray}
while the second one gives $\int_{-\infty}^0 dy e^{p y} \overline{G^-(x_0+y,  t|x_0 0)} =\rho^2 G^-(-p,p,\rho)$, i.e. the same function of $p$. Hence the final result is already symmetric. The Laplace inversion is thus immediate and one finds, using again (\ref{normal}):
\begin{eqnarray} 
&& \overline{G(x_0+y,t|x_0,0)} = (\frac{1 }{\bar \ell_\Gamma E^+_{\rho} E^-_\rho}   - \frac{1}{\rho}) g(|y|) e^{-\rho |y|} \label{GS1} \\
&&  = \overline{P(x_0,t|x_0,0)} \frac{1}{\Gamma^3} G(|y|/\Gamma^2) e^{- |y| (\rho+ \delta^2)}  \label{GS}
 \end{eqnarray}
 with $\Gamma= \ln t$, 
 using our previous result (\ref{return}) for the average return probability. At strictly coinciding points the two
 Green's function must coincide, see Eq. (\ref{relation}) both being then proportional to the same averaged trace over the same eigenenergies. Note however the divergence of $G(X) \sim 1/(\sqrt{\pi} X^{3/2})$ at small $X$ hence $g(|y|) \sim 1/(\sqrt{\pi} |y|^{3/2})$ (probability of degeneracies being related to the return probability of a Brownian motion). Hence the result (\ref{GS}) is meant to hold in the scaling region $y \sim \Gamma^2$ and to break down for $y = O(1)$ since the true function $g(y)$ should equal $g(y=0)=1$. At the end the result (\ref{GS}) is very similar to the one of \cite{sinaiRSRG} and 
 \cite{balentsfisher} in the case $\rho=0$, up to an exponential factor $e^{- \rho |y|}$, i.e. the absorbers act simply as an increase in the magnitude of the drift. 
 
\section{Conclusion} 
\label{sec10}

In summary we have computed various observables for the Sinai model in presence of a small density $\rho$ of perfect absorbers, using the strong disorder RSRG method. We have confirmed the result of TH for the power law decay in time of the average return probability, and for the density of states of the corresponding Fokker-Planck and Schrodinger operators. We found that in presence of a small drift the decay remains a power law with a larger exponent. We have also computed the diffusion front and described the crossover from smaller times where the diffusion scale grows as $L(t) \sim T \ln^2 t$ as in the absence of the absorbers, to the very large time limit when the survivors are found to be at a distance of order $\xi(t) = T \ln t/\sqrt{\rho}$ of their starting point, a new survival length scale. Remarkably their asymptotic distribution is a step function symmetric around the starting point, and even more remarkably it remains so in presence of a drift. Beyond this surviving zone around the starting point we found that the probability decays - in a scaling sense - also with a power law but with a larger exponent continuously increasing with the rescaled distance $x/\xi(t)$. A qualitatively similar behaviour was found in the meeting and survival probability of two particles. The statistics of eigenstates was also studied using the RSRG and the connection with the methods used by Texier was emphasized. It was found that keeping only the lowest eigenstate in each interval between scatterers leads to the same leading large time result. From there a saddle point argument was presented to clarify the mechanism for power law decay in this model, as resulting from the competition of large regions free of absorbers and their anomalously long decay time. Finally, we computed the average Green's function of the associated Schrodinger problem, which is found to exhibit the same spatial decay as in the absence of absorbers, up to a global $e^{-\rho |x|}$ factor.

There remains a host of quantities to compute for this problem using RSRG, as was done in Ref. \cite{sinaiRSRG}. It would also be interesting to develop other methods, such as the one used in \cite{ComtetDean,MajumdarComtet,sire2} to compute diffusion fronts and first passage probabilities. There also remains fundamental questions, and let us list a few. Since a Sinai walker equilibrated for a time $t=e^{\Gamma/T}$ in a valley visits a site at potential $U$ above the bottom of the valley 
proportionally to $t e^{-U/T}$, which is large for $U<\Gamma$, it is reasonable to expect that imperfect scatterers will renormalize to perfect ones. However 
a detailed study of this effect would be interesting. Other questions are to understand how general is the mechanism for the restoration of the symmetry in the presence of a drift observed here for some observables, what are the detailed properties of the eigenfunctions and the a.c. transport properties, whether the quenched versus annealed (in the probabilist sense) transition discussed in \cite{benarous} also occurs here, and whether this model and some of the phenomena arising here can be extended to higher dimension, or to quantum models. 

\bigskip

{\it Note added}: after the completion of this manuscript we were communicated unpublished notes by
A. Comtet, C. Texier and Y. Tourigny who obtained, via different methods and for a special value of the drift  in the biased case, a similar decay exponent for the return probability as obtained here. 

\bigskip 

{\it Acknowlegments}: I am grateful to C. Texier for a careful reading of the manuscript and useful remarks. I thank C. Hagendorf and A. Rosso for interesting discussions. 

\appendix

\section{Survival of two particles}

We may ask to compute the probability that two particles distant by $y_0>0$ be both still alive at time $t$. 
For simplicity we will compute the contribution, which we denote $S(y_0,t)$, to this probability which comes from the event when the two particles start in the same segment between absorbers. Hence this is a lower bound to the total probability. However, since when they start in different regions we can expect a faster decay, it already contains interesting information. 

Again, as in section \ref{meeting} both initial conditions must be in the bulk region in Fig. \ref{fig:1} each away from the edge bonds, i.e the absorbing zone. Hence it reads:
\begin{eqnarray}
S(y_0,t) = \rho^2 \int_0^\infty L dL e^{-\rho L} \int_0^L \frac{dx_0}{L} \left< \theta(\ell_1<x_0) \theta(x_0+y_0<L-\ell_{2k+2}) \right>_L
\end{eqnarray}
where $\left< .. \right>_L$ denotes the average with respect to the fixed $L$ measure (\ref{fs}). We have simply multiplied the probability that the leftmost particle happens to be on a segment of length $L$, with the
uniform probability $dx_0/L$ that it is at position $x_0$ within $dx_0$, and expressed the constraint. Introducing the Laplace transform w.r.t. $y_0$ and performing the integration $\int_0^{L-\ell_1-\ell_{2k+2}} dy_0 e^{-p y_0} \int_{\ell_1}^{L-\ell_{2k+2}-y_0} dx_0$ one obtains:
\begin{eqnarray}
&& S(p,t) = \int_0^\infty dy_0 e^{-p y_0} S(y_0,t) = \frac{1}{p^2} (\Phi(p) - p \Phi'(0) - \Phi(0)) \\
&& \!\!\!\!\!\! \!\!\!\!\!\!  \Phi(p) = \rho^2 \int_0^\infty dL e^{-\rho L} \left<  e^{- p(L-\ell_1-\ell_{2k+2})} \right>_L = \rho^2 \bar \ell_\Gamma \sum_{k=0}^\infty E^-_\Gamma(\rho) (P^+_\Gamma(p+\rho) P^-_\Gamma(p+\rho))^k E^+_\Gamma(\rho) \nonumber \\
&& = \rho^2  \frac{E^-_\Gamma(\rho)  E^+_\Gamma(\rho)}{(p+\rho) E^-_\Gamma(p+\rho)  E^+_\Gamma(p+\rho)} 
\end{eqnarray}
where $\Gamma=T \ln t$, and $S(y=0,t)=S(t)$ the single particle survival probability computed in
Section \ref{survivalproba}. 

In the symmetric case (up to an immaterial constant):
\begin{eqnarray}
&& \Phi(p) = \rho \tanh^2(\Gamma \sqrt{\rho}) \frac{1}{\sinh^2(\Gamma \sqrt{p+ \rho})}
\end{eqnarray}
Upon Laplace inversion one finds:
\begin{eqnarray}
&& \Phi(y) = \rho \tanh^2(\Gamma \sqrt{\rho}) \frac{1}{\Gamma^2} \sum_{m=-\infty}^{+\infty} (1 - 2 \pi^2 m^2 \frac{y}{\Gamma^2}) e^{- ( \frac{\pi^2}{\Gamma^2} m^2 + \rho) y } \\
&& = \rho \tanh^2(\Gamma \sqrt{\rho}) \frac{2 \Gamma}{y^{3/2} \sqrt{\pi}}  \sum_{n=-\infty}^{+\infty} n^2  e^{- n^2 \frac{\Gamma^2}{y} -  \rho y }  
\end{eqnarray}
We now have:
\begin{eqnarray}
&& S(y_0,t) = 2 \frac{\Gamma \sqrt{\rho}}{\sinh(2 \Gamma \sqrt{\rho})} - \rho y_0 \frac{1}{\cosh^2(\Gamma \sqrt{\rho})} 
+  \int_0^{y_0} dz \int_0^z dz' \Phi(z') \\
&& = \sum_{m=-\infty}^{+\infty}  \frac{\Gamma^4 \rho-\pi ^2 \Gamma^2 m^2 (2 \rho y_0+3)-2 \pi ^4 m^4 y_0}{\left(\Gamma^2 \rho+\pi ^2
   m^2\right)^3} e^{- ( \frac{\pi^2}{\Gamma^2} m^2 + \rho) y_0 } 
\end{eqnarray}

Through an analysis very similar to Section \ref{meeting} one finds the large time behaviour:
\begin{eqnarray}
&& S(y_0,t) \approx 4 T \ln t \sqrt{\rho} ( 1 - \frac{y_0 \sqrt{\rho}}{T \ln t})  t^{-2 T \sqrt{\rho}} 
\end{eqnarray}
for $0< y_0 < (T \ln t)/\sqrt{\rho}$, and a larger exponent for $y_0>T \ln t /\sqrt{\rho}$,
i.e. a decay $S(y_0,t) \sim t^{- T \sqrt{\rho} (\tilde y_0 + \frac{1}{\tilde y_0})}$ with $\tilde y_0=y_0 \sqrt{\rho}/(T \ln t)$.

\section{free diffusion}

For free diffusion (with unit diffusion coefficient) with absorbing boundaries on $[0,L]$ there are two dual formula for the diffusion front:
\begin{eqnarray}
&&  \!\!\!\!\!\! \!\!\!\!\!\! P_L(x t|x_0 0) = \frac{1}{\sqrt{4 \pi t}} \sum_{n=-\infty}^\infty ( e^{- \frac{(x-x_0 + 2 n L)^2}{4 t}} -  e^{- \frac{(x+x_0 + 2 n L)^2}{4 t}}) \theta(0<x<L) \theta(0<x_0<L)  \\
&& = \frac{2}{L} \sum_{m=1}^\infty \sin(\frac{\pi m x}{L}) \sin(\frac{\pi m x_0}{L}) e^{- \pi^2 m^2 t/L^2} \theta(0<x<L) \theta(0<x_0<L)
\end{eqnarray}

Considering now the problem of Poissonian absorbers, one averages over the distribution of interval sizes $L$ between absorbers and obtains the diffusion front averaged over (uniformly distributed) initial positions as:
\begin{eqnarray}
&& P(y,t)=  \rho^2 \int_0^\infty dL L e^{- \rho L} \theta(|y|<L) \int_0^{L-|y|} \frac{dx_0}{L} P_L(x_0+y, t|x_0 0) \\
&& =
\rho^2 \sum_{m=1}^\infty \int_{|y|}^\infty dL e^{- \rho L - \pi^2 m^2 t/L^2}  ( (1- \frac{|y|}{L}) \cos(\frac{\pi m |y|}{L}) 
+ \frac{1}{m \pi} \sin(\frac{\pi m |y|}{L}) ) 
\end{eqnarray}

From this expression, one obtains the average return to the origin probability, and its large time behavior:
\begin{eqnarray}
&& P(0,t) = \rho^2 \sum_{m=1}^\infty \int_0^\infty dL e^{- \rho L - \pi^2 m^2 t/L^2}  \approx \frac{2^{2/3} \pi ^{5/6}}{\sqrt{3}}  \rho^{4/3} t^{1/6} e^{-3 \left(\frac{\pi }{2}\right)^{2/3} \rho^{2/3} t^{1/3}}
\end{eqnarray}
retaining $m=1$ and using the saddle point at $L=2^{1/3} \pi^{2/3} t^{1/3} \rho^{-1/3}$ and computing the fluctuations around it. One also obtains the survival probability, and its large time behaviour as:
\begin{eqnarray}
&&  \!\!\!\!\!\! \!\!\!\!\!\! S(t) = \int_{-\infty}^\infty dy P(y,t) = \rho^2 \int_0^\infty dL L e^{- \rho L} \int_0^L \frac{dx_0}{L} \int_0^L dx
P_L(x,t|x_0,0) \\
&&  \!\!\!\!\!\! \!\!\!\!\!\! = \rho^2 \sum_{m=1}^\infty \int_0^\infty dL L \frac{4}{m^2 \pi^2} (1-(-1)^m)  e^{- \rho L - \pi^2 m^2 t/L^2}  \approx \frac{16}{\sqrt{3 \pi}} \rho t^{1/2} e^{-3 \left(\frac{\pi }{2}\right)^{2/3} \rho^{2/3} t^{1/3}}
\end{eqnarray}
using again the saddle point and $m=1$. At large time the full diffusion front is also dominated by $m=1$:
\begin{eqnarray}
&& P(y,t) \approx y \rho^2 \int_1^\infty d\tilde L e^{- \rho y \tilde L - \pi^2 t/(y^2 \tilde L^2)}  ( (1- \frac{1}{\tilde L}) \cos(\frac{\pi}{\tilde L}) 
+ \frac{1}{\pi} \sin(\frac{\pi}{\tilde L}) )
\end{eqnarray}
with $\tilde L=L/y$. Upon a similar saddle point analysis one finds that the diffusion front expressed
in the variable:
\begin{eqnarray}
&& \tilde y = y/\xi_0(t) \quad , \quad \xi_0(t) = 2^{1/3} \pi^{2/3} \rho^{-1/3} t^{1/3}
\end{eqnarray}
takes a simple scaling form:
\begin{eqnarray}
&& P(y,t) dy = S(t)  \frac{\pi^2}{8} \big((1- |\tilde y|) \cos(\pi |\tilde y|) + \frac{1}{\pi} \sin(\pi |\tilde y|) \big) \theta(1-|\tilde y|) d \tilde y
\end{eqnarray}
Hence the survival zone has also sharp edges at $\tilde y=\pm 1$, though this time it is continuous at the edge (the rescaled probability vanishes as $ \sim (1-|\tilde y|)^3$ at the edge). By the same mechanism as described in the text, the decay of the average front {\it outside} this region is as $\exp(- C(\tilde y) t^{1/3})$, with $C(\tilde y)=\frac{1}{2} C(1) (\tilde y + \frac{1}{\tilde y^2}) $, i.e. the amplitude in the stretched exponential grows continuously with $\tilde y>1$.




\section*{References}

\begin{thebibliography}{999}





\bibitem{sinai} Ya. G. Sinai, Theory of Prob. and Appl. 27(2), 247 (1982).

\bibitem{oldresults} J.-P. Bouchaud, A. Comtet, A. Georges, and P. Le Doussal, Europhys. Lett. 3 653 (1987) and Ann. Phys. (N.Y.) 201, 285Ð341 (1990). 

\bibitem{biophysics}
David K. Lubensky, David R. Nelson, arXiv:cond-mat/0004423, 
Phys. Rev. Lett. 85, 1572 (2000) and arXiv:cond-mat/0107423. 
D. R. Nelson, arXiv:cond-mat/0309559. 

\bibitem{vinokur}
I. Aranson, L.Tsimring, V. Vinokur, adap-org/9702002. 

\bibitem{finance}
D. Dufresne, Scand. Act. J. (1990) 39, H. Geman and M. Yor, Math. Fin. 3 (1993) 349.


\bibitem{pastur} I. M. Lifshits, S. A. Gredeskul, and L. A. Pastur, Introduction to the theory of disordered systems, John Wiley and 
Sons, 1988, A. A. Gogolin and V. I. MelÕnikov, Sov. Phys. JETP 
46, 369 (1977). A. A. Gogolin, Phys. Rep. 86(1), 1Ð53 (1982)


\bibitem{texierdos}
C. Texier, J. Phys. A: Math. Gen. 33, 6095 (2000).


\bibitem{balentsfisher} 
L. Balents, M. P. A. Fisher, arXiv:cond-mat/9706069.


\bibitem{Kesten}
H. Kesten, M. Koslov and F. Spitzer, Compositio Math. 30 (1975) 145, H. Kesten, Physica 138 A (1986) 299.

\bibitem{golosov} A.O. Golosov, Comm. Math. Phys. 92 (1984) 491

\bibitem{sinaiRSRG}
P. Le Doussal, C. Monthus, and D. S. Fisher, condmat/9811300, Phys. Rev. E 59(5), 4795 (1999). 
D. Fisher, P. Le Doussal and C. Monthus, Phys. Rev. Lett. 80 (1998) 3539.

\bibitem{fisher1}
D. S. Fisher Phys. Rev. B 50, 3799 (1994) and 
D. S. Fisher Phys. Rev. B 51, 6411-6461 (1995).

\bibitem{igloimonthus}
For a general review see:
F. Igloi, C. Monthus, arXiv:cond-mat/0502448, Physics Reports 412, 277-431, (2005)


\bibitem{HT2}
C. Texier and C. Hagendorf  arXiv:0902.2698. 


\bibitem{HT1}
C. Hagendorf and C. Texier, J. Phys. A: 
Math. Theor. 41, 405302 (2008). 


\bibitem{fisheryoung} 
D.S. Fisher and A. P. Young, Phys. Rev. B 58, 9131 (1998)

\bibitem{monthusFS}
C. Monthus, arXiv:cond-mat/0309029, Phys. Rev. B 69, 054431 (2004).

\bibitem{cecile}
C. Monthus, arXiv:cond-mat/0212212, Phys. Rev. E 67, 046109 (2003). 


\bibitem{energyRSRG}
see Sec. VIII-A in C. Monthus and P. Le Doussal, arXiv:cond-mat/0202295, Phys. Rev. E 65 (2002) 66129.

\bibitem{distributions} 
A. Comtet, C. Monthus, M. Yor, arXiv:cond-mat/9601014, J. Appl. Proba. 35 (1998) 255. 

\bibitem{footnote1}
see e.g. Ref. \cite{oldresults,sinaiRSRG} for discussion of this model-dependent crossover at very short times, not studied here.

\bibitem{footnote2}
at larger $\rho$, it remains to be understood if the coincidence of prefactor with TH is accidental or not. 

\bibitem{footnote3} this amounts to neglect relaxation on times scales $t_j=1/E_j$ w.r.t. $t_i$ with $j>i$. This is the standard argument of RSRG being asymptotically exact: this neglect can only produce an uncertainty in $\Gamma_i$ of order $T \ln (t_i \pm t_j) = T \ln t_i + \ln(1 \pm e^{- (\Gamma_i - \Gamma_j)/T})=\Gamma_i + O(e^{-(\Gamma_i - \Gamma_j)/T})$. For decimations which occur in the same region of space (i.e. overlapping eigenfunctions) $\Gamma_i - \Gamma_j $ is of the same order as $\Gamma_i$ hence the error is exponentially small in $\Gamma$ at large $\Gamma$.

\bibitem{sire1}
D. S. Dean, C. Sire, J. Sopik, arXiv:cond-mat/0604456, Phys. Rev. E 73, 066130 (2006)


\bibitem{greginprep}
P. Le Doussal and G. Schehr, in preparation. 

\bibitem{ComtetDean}
D. Dean and A. Comtet, J. Phys. A31 (1998) 8595

\bibitem{MajumdarComtet}
S. Majumdar and A. Comtet, Phys. Rev. E {\bf 66} (2002) 061105.

\bibitem{sire2}
C. Sire, arXiv:cond-mat/9902223, Phys. Rev. E 60, 1464 (1999).


\bibitem{benarous}
G.Ben Arous, S. Molchanov
and A. F. Ramirez, arXiv:math/0501107v3,
The Annals of Probability
2005, Vol. 33, No. 6, 2149. 


\end{thebibliography}
\end{document}